\documentclass[aps,prl,review,groupedaddress]{revtex4-1}
\usepackage{natbib}

\usepackage{indentfirst}
\usepackage{bm}
\usepackage{array, threeparttable}
\usepackage{multirow}
\usepackage{mathrsfs}
\usepackage{amsmath, amssymb}

\usepackage{graphicx, subfigure}
\begin{document}


\title{Implementation of dual time stepping and GMRES of implicit gas-kinetic scheme for unsteady flow simulations}


\author{Ji Li}
\email[]{leejearl@mail.nwpu.edu.cn}
\affiliation{National Key Laboratory of Science and Technology on Aerodynamic Design and Research, Northwestern Polytechnical University, Xi'an, Shaanxi 710072, China}

\author{Chengwen Zhong}
\email[]{Corresponding author: zhongcw@nwpu.edu.cn}
\affiliation{National Key Laboratory of Science and Technology on Aerodynamic Design and Research, Northwestern Polytechnical University, Xi'an, Shaanxi 710072, China}

\author{Yong Wang}
\affiliation{National Key Laboratory of Science and Technology on Aerodynamic Design and Research, Northwestern Polytechnical University, Xi'an, Shaanxi 710072, China}

\author{Congshan Zhuo}
\affiliation{National Key Laboratory of Science and Technology on Aerodynamic Design and Research, Northwestern Polytechnical University, Xi'an, Shaanxi 710072, China}

\date{\today}

\begin{abstract}
 A dual time-stepping strategy of gas-kinetic scheme for the simulation of unsteady flows is introduced in this work. The dual time-stepping strategy is always used in the approaches of unsteady flows, and the ability of dual time-stepping to accelerate the computation with acceptable error tolerance is evident. In our paper, we adopt the techniques of dual time-stepping methods for implicit gas-kinetic scheme to simulate the unsteady flows, which is very popular for the numerical methods based on the Navier-Stokes equations. It is carried out by (a) solving the gas-kinetic scheme in finite volume method; (b) obtaining the inviscid flux Jacobian by Roe scheme; (c) involving the computation of viscous flux Jacobian which is not mentioned in the previous implicit gas-kinetic schemes; (d) approximating the linear system of pseudo steady state by generalized minimal residual algorithm (GMRES). The explicit gas-kinetic scheme has been proved to be an accurate approach for both the steady and unsteady flows, and the implicit gas-kinetic scheme is also be developed to accelerate the convergence of steady state. The dual time-stepping method proposed in our study is of great importance to the computations of unsteady flows. Several numerical cases are performed to evaluate the behavior of dual time-stepping strategy of gas-kinetic scheme. The incompressible flow around blunt bodies (stationary circular cylinder and square cylinder) and the transonic buffet on the NACA0012 airfoil are simulated to demonstrate the overall performance of the proposed method which is applicable to the fluid flows from laminar to turbulent and from incompressible to compressible.
\end{abstract}


\maketitle

\clearpage

\section{Introduction}
Gas-kinetic scheme developed by Xu\cite{xu2001gas}\cite{xu2005multidimensional} is a unified solver of computational fluid dynamics (CFD) for both incompressible and compressible flows. Base on the
Bhatnagar-Gross-Krook\cite{bhatnagar1954model} (BGK) model, gas-kinetic scheme describes the macroscopic fluid flows by microscopic distribution functions. Many published results have demonstrated the accuracy and efficiency of gas-kinetic scheme in the simulations of laminar\cite{xu2001gas}\cite{xu2005multidimensional}\cite{yuan2015immersed} and turbulent flows\cite{li2010numerical}\cite{xiong2011numerical}\cite{righi2014modified}\cite{righi2016gas}\cite{ong2014bgk}\cite{FLD:FLD4239}.

The approach of unsteady flows become more and more important in the field of engineering. Explicit scheme can be seen as the best choice for the simulation of unsteady flows with sufficient accuracy. But, the physical time scales might be much more large than the explicit time step which is determined by the CFL numbers in some cases, and this will lead to very expensive computational cost. Coupled with the extremely complex computation of flux in gas-kinetic scheme, it is highly necessary to develop a fast algorithm for the gas-kinetic scheme to simulate the unsteady flows.

Dual time-stepping strategy is one of the popular methods used widely and has been proved as an effective method for unsteady flows without impairing the accuracy. To accelerate the computation of unsteady flows, inner iteration should be employed in each physical time step for the convergence of pseudo steady state. The solutions of pseudo steady state can be obtained by local time-stepping, multigrid and implicit scheme\cite{jameson1991time}\cite{pulliam1993time}\cite{blazek2015computational}. In our paper, the implicit gas-kinetic scheme and the local time-stepping are used to approach the pseudo steady state. O. Chit\cite{chit2004implicit} proposed an implicit gas-kinetic method based on the approximate factorization-alternating direction implicit (AF-ADI) scheme, and the results give a good agreement with the compared data. K. Xu and M. Mao\cite{xu2005multidimensional} developed an implicit scheme based on the Euler fluxes and LU-SGS method, and the method is applied to simulate hypersonic laminar viscous flows.  J. Jiang and Y. Qian\cite{jiang2012implicit} make a comparison of implicit GKS and the multigrid GKS in 3D simulations. W. Li\cite{li2014implicit} proposed an unstructured implicit GKS based on the LU-SGS method. In our paper, we adopt the generalized minimal residual method (GMRES)\cite{saad1986gmres}\cite{saad2003iterative}\cite{van2003iterative} method into gas-kinetic scheme to solve the linear systems of flux Jacobian matrix, and the linear system is constructed not only by using the Euler flux Jacobian, but also by using the viscous flux Jacobian which is not mentioned in the previous implicit gas-kinetic schemes.

In our study, we set up three test cases to validate the dual time-stepping strategy for gas-kinetic scheme, and several features of the free stream flow conditions have been considered. The case of incompressible flow around the circular cylinder is focus on the simulation of unsteady flows with the low Reynolds number, which also has been implemented by Yuan\cite{yuan2015immersed} using gas-kinetic scheme with immersed boundary method. The effect of dual time-stepping method on the incompressible turbulent flow is demonstrated in the second test case. The vortex shedding frequency, surface loads of this flows are obtained. The last case is about the transonic buffet on the NACA0012 airfoil, which can be referred to many articles, such as McDevitt and Okuno\cite{mcdevitt1985static}, J. Xiong\cite{xiong2012computation}, M. Iovnovich\cite{iovnovich2012reynolds}, C.Q. Gao\cite{gao2015numerical}.  For the approach of turbulent flows, turbulence models\cite{spalart1992one}\cite{menter1994two} are coupled with gas-kinetic scheme. All the tests set up in our study obtain a good accordance with the experiments and other numerical methods (Since, there are only very few literatures of gas-kinetic schemes focused on the simulation of unsteady flows, most of the methods are based on the Navier-Stokes equations).

The rest of our paper is organized as follows. In the second section, the gas-kinetic scheme, the dual time-stepping strategy, the flux Jacobian and the GMRES method are introduced briefly. In the third section, three numerical test cases (incompressible laminar flow over the stationary circular cylinder, incompressible turbulent flow around a square cylinder, and the transonic buffet on the NACA0012 airfoil surface with high Reynolds number) are conducted for different purposes. Finally, a short conclusion is summarized in the final section.

\section{Numerical Methods}\label{NumericalMethods}

\subsection{Gas-kinetic scheme}
In this section, the procedure of gas-kinetic scheme proposed by Xu\cite{xu2001gas} is introduced briefly.

\subsubsection{Initial reconstruction}
Similar to other finite volume methods, gas-kinetic scheme in finite volume method can be expressed as
\begin{equation}\label{gks-fvm}
    \bm{w}_I^{n+1}=\bm{w}_I^{n} - \frac{1}{\Omega_I} \int_{t_n}^{t_{n+1}} \sum_{J=1}^{N_{IF}}\bm{F}(t)_JS_Jdt,	
\end{equation}
where $I$ is the index of the finite volumes, $J$ means the index of interface belonged to the cell $I$, $N_{IF}$ is the
total number of the cell interfaces around the finite volume $I$, $\Omega_I$ denotes the measure of the finite volume $I$,
$\bm{F}$ represents the flux across the cell interface, and $S_J$ is the measure of the $Jth$ cell interface.  The
macroscopic variable $\bm{w}$ appeared in the Eq.~\ref{gks-fvm} reads as
\begin{equation}\label{macroVar}
	\bm{w} = \left(
		\begin{array}{c}
			\rho \\
			\rho U_i \\
			E
		\end{array}
\right) = \int {\bm{\psi}fd\Xi},\quad \bm{\psi} = \left(1, u_i, \frac{1}{2}\left(\bm{u} \cdot \bm{u} + \bm{\xi} \cdot \bm{\xi} \right) \right)^T,
\end{equation}
and the flux $\bm{F}$ at the cell interface is
\begin{equation}\label{flux}
	\bm{F} = \left(F_\rho, F_{\rho U_i}, F_{\rho e} \right)^T = \int {(\bm{u} \cdot \bm{n})\bm{\psi}fd\Xi},
\end{equation}
where $f$ is the distribution function and $d\Xi = \left(\prod\limits_{i=1}^{D}du_i\right)
\left(\prod\limits_{i=1}^{K}d\xi_i \right)$, $D$ represents the dimension, $K$ denotes the total degree of freedom of
internal variables, $\bm{\xi}$ means the internal variables, $\rho$ is the density, $\bm{U}$ is the macroscopic velocity and $E$ is the energy of gas in the finite volume, $\bm{u}$ is the particle velocity, and $\bm{n}$ represents the normal vector pointing outside of the finite volume respectively.

For a finite volume method, flux across the cell interface is based on initial reconstruction in which interpolation
techniques and limiters are used. For unstructured grids, it is proved that the Venkatakrishnan\cite{venkatakrishnan1995convergence}
limiter works, which has been used in our paper.

The conservative variable in the finite volume can be expressed as
\begin{equation}\label{reconstruction1}
	\bm{w}_I(\bm{x}) = \bm{w}_I(\bm{x}_I) + ((\bm{x}-\bm{x}_I) \cdot \nabla \bm{w}_I)\bm{\phi}_I,
\end{equation}
where $\bm{\phi}_I$ denotes the limiter in the finite volume $I$, $\bm{x}_I$ is coordinate of the cell center, $\bm{x}$ means the position of a point located in the finite volume, $\bm{w}_I$ represents the average conservative variable of the finite volume $I$, and $\nabla \bm{w}_I$ denotes the spatial gradient of conservative variable in the finite volume.

The Venkatakrishnan limiter employed in our study reads as
\begin{equation}\label{VenLimiter}
	\bm{\phi}_{IJ}=\left \{
		\begin{array}{ll}
			L(\bm{w}_I^M-\bm{w}_I, \Delta_{IJ}), & \Delta_{IJ} > 0 \\
			L(\bm{w}_I^m-\bm{w}_I, \Delta_{IJ}), & \Delta_{IJ} < 0 \\
			1, & \Delta_{IJ} = 0
		\end{array}
	\right.
\end{equation}
where
\begin{equation}\label{venL}
	L(a,b)=\frac{a^2+2ab+\epsilon}{a^2+2b^2+ab+\epsilon}
\end{equation}
\begin{equation}\label{phiI}
	\bm{\phi}_I=\min {\bm{\phi}_{IJ}}
\end{equation}
\begin{equation}\label{DeltaIJ}
	\Delta_{IJ} = (\bm{x}_J-\bm{x_I}) \cdot \nabla \bm{w}_I
\end{equation}
\begin{equation}\label{epsilonV}
    \epsilon = (\zeta \bar{h})^3, \zeta > 0.
\end{equation}
$J$ represents the index of cell interface surrounding the finite volume $I$, $\bar{h}$ denotes the average cell size of the grid in the computational domain, $\zeta$ is a constant number, $\bm{w}_I^M$ and $\bm{w}_I^m$ are the maximum and minimum values of macroscopic conservative variables in the neighbors of the cell $I$ respectively. The value of $\zeta$ has a great effect on the accuracy and convergence of the numerical algorithm. Because of determining a proper value of $\zeta$ is a confused and difficult problem in practice, it is hard to get a suitable value for $\zeta$. In our work, we follow the ideas in Ref. \cite{wang2000fast} which modified Eq.~\ref{epsilonV} as

\begin{equation}\label{mepsilon}	
	\epsilon = \eta(\bm{w}^{Max} -\bm{w}^{min}), \eta\in (0.01, 0.2).	
\end{equation}
Since $\bm{w}^{Max}$ and $\bm{w}^{min}$ are the maximum and minimum values of macroscopic conservative variables in the whole computational domain, which do not rely on the local value and provide a threshold value for the smooth region. In our tests, the value of $\eta$ is given as $0.15$ following the suggestion in Ref. \cite{li2016gas}.

\subsubsection{Flux across the cell interface}
After the reconstruction stage, $\bm{F}$ can be obtained using Eq.~\ref{flux}. Up to now, the only issue is to calculate the distribution function $f$ at the cell interface. In this paper, we take a special case, in which the interface is normal to x-axis, to demonstrate the computing procedure of the flux at the cell interface. In practice, the cell interface is rarely normal to the x-axis, especially for grids with triangles and tetrahedrons. So, the transformation of coordinate system must be applied. $f$ can be written as
\begin{equation}\label{finterface}
	\begin{array}{ccc}
		f & = & (1-e^{-\frac{t}{\tau}})(g_0-t\bm{u} \cdot \nabla{g_0}) + e^{-\frac{t}{\tau}}(\bar{g}-t\bm{u} \cdot \nabla{\bar{g}})
		+t(\frac{\partial g_0}{\partial{t}}+\bm{u} \cdot \nabla{g_0}) \\
		&   & -\tau(1-e^{-\frac{t}{\tau}})(\frac{\partial g_0}{\partial{t}}+\bm{u} \cdot \nabla{g_0}) - \tau e^{-\frac{t}{\tau}}(\frac{\partial{\bar{g}}}{\partial{t}}+\bm{u} \cdot \nabla{\bar{g}}). \\
	\end{array}
\end{equation}
For notational convenience we define
\begin{equation}\label{non-eq2}
		\bar{g} = (1-H(\bm{x}\cdot \bm{n}))g^l + H(\bm{x}\cdot \bm{n})g^r,
\end{equation}
where $H(x)$ is the Heaviside function
\begin{equation}\label{heaviside}
	H(x) = \left \{
		\begin{array}{cc}
			0, & x < 0, \\
			1, & x \geq 0.
		\end{array}
	\right.
\end{equation}
where $g_0$, $g^l$ and $g^r$, obtained after the initial reconstruction of macroscopic conservative variables, are the Maxwellian distribution function at and both sides of cell interface respectively. $\nabla g_0$, $\nabla g^l$, and $\nabla g^r$ are the spatial gradients of $g_0$, $g^l$ and $g^r$. $\frac{\partial g_0}{\partial t}$, $\frac{\partial g^l}{\partial t}$, and $\frac{\partial g^r}{\partial t}$ are the time derivatives. The kernel of gas-kinetic scheme is to compute the distribution function $f$ at the cell interface, and the detailed determinations of $\nabla g_0$, $\nabla g^l$, $\nabla g^r$, $\frac{\partial g_0}{\partial t}$, $\frac{\partial g^l}{\partial t}$, and $\frac{\partial g^r}{\partial t}$ can be seen in Ref.  \cite{xu2001gas}\cite{may2007improved}.

\subsubsection{Collision time and numerical viscosity}
The collision time appeared in Eq.~\ref{finterface} is defined as
\begin{equation}\label{tau}
    \tau = \frac{\mu}{p} + \frac{|p_l-p_r|}{|p_l+p_r|}\Delta t,
\end{equation}
where $p$ is the pressure, $\mu$ is the dynamic viscosity coefficient and satisfying Sutherland's Law
\begin{equation}\label{Sutherland}
    \frac{\mu}{\mu_{ref}} = \left(\frac{T}{T_{ref}} \right)^{\frac{3}{2}}\frac{T_{ref}+S}{T+S},
\end{equation}
where $\mu_{ref} = 1.7894 \times 10^{-5}kg/(m \cdot s)$, $T_{ref} = 288.16K$, $S = 110K$, and $T$ is the temperature
corresponding with $\mu$.  The second term on the right hand side of Eq.~\ref{tau} represents the artificial numerical viscosity. $\Delta t$ is the explicit time step, which can be calculated by
\begin{equation}\label{exDeltaT}
    \Delta t = CFL \cdot \min \left\{ \frac{{\Omega}_I}{({\varLambda}_c)_I} \right\}.
\end{equation}
Where ${\varLambda}_c$ reads as
\begin{equation}\label{varLambdac}
    {\varLambda}_c = \sum_{J=1}^{N_{IF}}{\left(\left(\lvert \bm{u}\cdot \bm{n}_J\rvert + a_s\right)S_J\right)}.
\end{equation}
$N_{IF}$ in Eq.~\ref{varLambdac} represents the number of cell interfaces around the finite volume $I$, $\bm{u}$ denotes the macroscopic (averaged) velocity in the finite volume, $S_J$ means the measure of $Jth$ cell interface, and $a_s$ represents the sound speed in the finite volume.

For the prediction of turbulent flows, Eq.~\ref{tau} can be rewritten as
\begin{equation}\label{modified-tau}
    \tau = \frac{\mu + \mu_t}{p} + \frac{|p_l-p_r|}{|p_l+p_r|}\Delta t,
\end{equation}
where $\mu_t$ is the turbulent eddy viscosity, and it comes from the allied turbulence model. There are other
techniques, Chen et al.\cite{chen2003extended} and Succi et al.\cite{succi2002towards}, used for modifying the collision time, and we use Eq.~\ref{modified-tau} in our paper for simplicity.

\subsection{Dual time stepping strategy}\label{secDualTimeStepping}
The simulation of unsteady flows phenomena is of more and more importance in many disciplines of engineering. Explicit scheme is considered as the best choice for the simulation of unsteady flows with great accuracy. But, in some cases, such as the unsteady turbulent flows, the physical time scales might be very large in comparison to the explicit time steps which are determined by CFL numbers. Since predicting such flows using explicit scheme spends so long times, computational costs are very expensive. It is necessary to develop less expensive methods  without impairing the accuracy of the prediction. In this section, dual time-stepping strategy, which is very popular for unsteady flows, is introduced.

The explicit and implicit schemes can be expressed as one basic non-linear schemes. It reads as
\begin{equation}\label{basicscheme}
    \frac{\Omega_I}{\Delta t_I}\Delta \bm{w}_I^n=-\frac{\beta}{1+\omega}\bm{R}_I^{n+1}
        - \frac{1-\beta}{1+\omega}\bm{R}_I^n
        + \frac{\omega}{1+\omega}\frac{\Omega_I}{\Delta t_I}\Delta\bm{w}_I^{n-1},
\end{equation}
where $\Delta \bm{w}_I^n=\bm{w}_I^{n+1}-\bm{w}_I^n$, $\Delta t_I$ represents the local time step, and the parameters $\beta$ and $\omega$ appeared in Eq.~\ref{basicscheme} are used to determine the type (explicit or implicit) and also the temporal accuracy.

Dual time-stepping strategy is based on the Eq.~\ref{basicscheme}. We set $\beta = 1$ and $\omega = 0.5$. Hence, we obtain
\begin{equation}\label{dual1}
    \frac{3\Omega_I^{n+1}\bm{w}_I^{n+1}-4\Omega_I^n\bm{w}_I^n+\Omega_I^{n-1}\bm{w}_I^{n-1}}{2\Delta t_p}=-\bm{R}_I^{n+1},
\end{equation}
where $\Delta t_p$ denotes the global physical time step and Eq.~\ref{dual1} is a second order time accurate version of Eq.~\ref{basicscheme}.  The left side of Eq.~\ref{dual1} is a three-point backward-difference approximation of the time derivation. Thus, Eq.~\ref{dual1} can be treated as a modified steady state problem to be solved using a pseudo-time step $t^*$
\begin{equation}\label{dual2}
    \frac{\partial (\Omega_I^{n+1}\bm{w}_I^{*})}{\partial t^*} = - \bm{R}_I^*(\bm{w}_I^*),
\end{equation}
where $\bm{w^*}$ is the approximation to $\bm{w^{n+1}}$. The unsteady residual can be expressed as
\begin{equation}\label{unsteadyRes}
    \bm{R}_I^*(\bm{w}_I^*) = \bm{R}(\bm{w}_I^*) + \frac{3}{2\Delta t_p}(\Omega_I^{n+1}\bm{w}_I^*)-\bm{Q}_I^*,
\end{equation}
where $\bm{Q}_I^*$ represents the source term,
\begin{equation}\label{dualSource}
    \bm{Q}_I^* = \frac{2}{\Delta t_p} \Omega_I^n\bm{w}_I^n-\frac{1}{2\Delta t_p}\Omega_I^{n-1}\bm{w}_I^{n-1}.
\end{equation}

The steady state solution of Eq.~\ref{dual2}, which is solved using GMRES method in our paper, approximates the
macroscopic flow variables at the time step level $n+1$, i.e., $\bm{w}^* = \bm{w}^{n+1}$. To apply an implicit scheme for the steady solution $\bm{w}^*$ in pseudo time $t^*$, the first stage is to formulate Eq.~\ref{dual2}
as an nonlinear implicit scheme as follow
\begin{equation}\label{dualimplicit1}
    \frac{\partial \bm{w}_I^*}{\partial t^*} = - (\bm{R}_I^*)^{l+1},
\end{equation}
where $l+1$ is the new time level of pseudo-time. Then, the right side of Eq.~\ref{dualimplicit1} can be linearised as
\begin{equation}\label{linearUnsteadyRes}
    (\bm{R}_I^*)^{l+1} \approx (\bm{R}_I^*)^{l} + \frac{\partial \bm{R}_I^*}{\partial \bm{w}^*}\Delta \bm{w}_I^*,
\end{equation}
where
\begin{equation}
    \Delta \bm{w}^* = (\bm{w}^*)^{l+1} - (\bm{w}^*)^l,
\end{equation}
\begin{equation}\label{dualJac}
    \frac{\partial \bm{R}_I^*}{\partial \bm{w}^*} = \frac{\partial \bm{R}_I}{\partial \bm{w}} + \frac{3\Omega}{2\Delta t_p}.
\end{equation}
Substituting Eq.~\ref{linearUnsteadyRes} and Eq.~\ref{dualJac} into Eq.~\ref{dualimplicit1}, we get the following implicit scheme
\begin{equation}\label{dualimplicit2}
    \left [ \left( \frac{1}{\Delta t^*} + \frac{3}{2\Delta t_p}\right)\Omega_I^{n+1}
    + \left( \frac{\partial \bm{R}}{\partial \bm{w}}\right)_I\right]\Delta \bm{w}_I^* = -(\bm{R}^*)^l.
\end{equation}
Let
\begin{equation}\label{linearSysA}
\bm{A} = \left [ \left( \frac{1}{\Delta t^*} + \frac{3}{2\Delta t_p}\right)\Omega_I^{n+1}
+ \left( \frac{\partial \bm{R}}{\partial \bm{w}}\right)_I\right],
\end{equation}
\begin{equation}\label{linearSysX}
\bm{X} =\Delta \bm{w}_I^*,
\end{equation}
\begin{equation}\label{linearSysB}
\bm{B} = -(\bm{R}_I^*)^l,
\end{equation}
Eq.~\ref{dualimplicit2} can be rewritten as
\begin{equation}\label{AXB}
\bm{A}\bm{X}=\bm{B}.
\end{equation}
For solving the linear system of Eq.~\ref{AXB}, we employ the GMRES method in our paper.

\subsection{Flux Jacobian}
In the subsection \ref{secDualTimeStepping}, a linear system Eq.~\ref{AXB} is constructed for the implicit gas-kinetic scheme, and both the implements of implicit gas-kinetic scheme in structured grids\cite{xu2005multidimensional} and unstructured grids\cite{li2014implicit} have been developed by other researchers. In this section, we only focus on the determination of flux Jacobian at the cell interface.

\begin{figure}[!htp]
	\centering
	\includegraphics[width=0.2 \textwidth]{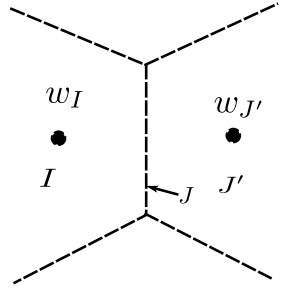}
	\caption{\label{figFluxJac1} The finite volumes at both sides of the $Jth$ interface.}
\end{figure}

In order to employ the implicit gas-kinetic scheme, the time averaged flux is needed. For a gas-kinetic scheme, the time averaged flux function reads as
\begin{equation}\label{timeAvgFlux}
    \overline{\bm{F}}_J = \frac{1}{\Delta t}\int_{t_n}^{t_n+\Delta t} \bm{F}(t)_JS_Jdt.
\end{equation}
where $\Delta t$ means the explicit time step determined by Eq.~\ref{exDeltaT}. $\bm{R}_I^{n+1}$ in the right side of Eq.~\ref{dual1} can be written as
\begin{equation}\label{Rn1}
    \bm{R}_I^{n+1} = \frac{1}{\Delta t} \int_{t_n}^{t_n+\Delta t} \sum_{J=1}^{N_{IF}}\bm{F}(t)_JS_Jdt	
    = \sum_{J=1}^{N_{IF}}\left(\frac{1}{\Delta t} \int_{t_n}^{t_n+\Delta t} \bm{F}(t)_JS_Jdt\right),	
\end{equation}
and then,
\begin{equation}\label{Rn2}
    \bm{R}_I^{n+1} = \sum_{J=1}^{N_{IF}}\overline{\bm{F}}_J.	
\end{equation}
Thus,
\begin{equation}\label{fluxJacobian1}
    \frac{\partial \bm{R}_I^{n+1}}{\partial \bm{w}_I}\Delta \bm{w}_I^n
    = \sum_{J=1}^{N_{IF}}\left(\frac{\partial \overline{\bm{F}}_J}{\partial \bm{w}_J}\Delta \bm{w}_J^{n}\right),	
\end{equation}
where $\Delta \bm{w}^{n} = \bm{w}^{n+1} - \bm{w}^n$.  Although the expression of flux Jacobian has been given in Eq.~\ref{fluxJacobian1}, it is still difficult to be computed based on the BGK model. In our study, we construct the flux Jacobian based on the Euler equations and Navier-Stokes equations. The partial derivative in the right and left side of Eq.~\ref{fluxJacobian1} can be decomposed as
\begin{equation}\label{fluxJacobian2}
    \left(\frac{\partial {\bm{R}}}{\partial \bm{w}}\right)_I = \left(\frac{\partial {\bm{R}_c}}{\partial \bm{w}}\right)_I
    +\left(\frac{\partial {\bm{R}_v}}{\partial \bm{w}}\right)_I
\end{equation}
and
\begin{equation}\label{fluxJacobian3}
    \left(\frac{\partial \overline{\bm{F}}}{\partial \bm{w}}\right)_J
    = \left(\frac{\partial \overline{\bm{F}_c}}{\partial \bm{w}}\right)_J
    +\left(\frac{\partial \overline{\bm{F}_v}}{\partial \bm{w}}\right)_J
\end{equation}
respectively.  Where $\bm{R}_c$ and $\overline{\bm{F}_c}$ are corresponding to the convective part. $R_v$ and
$\overline{\bm{F}_v}$ are corresponding to the viscous part.  For the convective part, we employ the flux Jacobian due to Roe scheme\cite{venkatakrishnan1991preconditioned} as follows
\begin{equation}\label{fluxJacRoe1}
    \frac{\partial \bm{R}_{cI}}{\partial \bm{w}}\Delta \bm{w}_I^n = \frac{1}{2}\sum_{J=1}^{N_{IF}}\left(
    \left(\frac{\partial \overline{\bm{F}_c}}{\partial \bm{w}}\Delta \bm{w}^n\right)_I
    +\left(\frac{\partial \overline{\bm{F}_c}}{\partial \bm{w}}\Delta \bm{w}^n\right)_{J'}
    -\lvert \overline{\lambda}_{Roe}\rvert_J(\bm{w}_{J'}^n-\bm{w}_I^n)
    \right),
\end{equation}
and Fig.~\ref{figFluxJac1} plots the finite volumes at both sides of the $Jth$ interface.

The viscous part is very important for the simulation of viscous flows, but it is not yet mentioned in the previous
implicit gas-kinetic schemes. In our study, it can be written as
\begin{equation}\label{fluxJacRoe2}
    \left(\frac{\partial \overline{\bm{F}}_v}{\partial \bm{w}}\right)_J\Delta \bm{w}_J^n
    \approx \left(\frac{\partial \overline{\bm{F}}_v}{\partial \bm{w}}\right)_{J'}\Delta \bm{w}_{J'}^n
    -\left(\frac{\partial \overline{\bm{F}}_v}{\partial \bm{w}} \right)_I\Delta \bm{w}_{I}^n.
\end{equation}
The details of $\frac{\partial \bm{F}_c}{\partial \bm{w}}$, $\frac{\partial \bm{F}_c}{\partial \bm{w}}$, Eq.
\ref{fluxJacRoe1} and Eq.~\ref{fluxJacRoe2} can be seen in the literature\cite{blazek2015computational}, and up to now, the computation of flux Jacobian is completed.

\subsection{GMRES method}
Implicit schemes used to accelerate the convergence behaviors are always resulted in the solving of linear systems like Eq.~\ref{AXB}.  The GMRES Method, originally suggested by Saad and Schulz\cite{saad1986gmres}, is one of the popular method used widely.

Defined $\bm{A}$ as an $n\times n$ real matrix and $\mathcal{K}$ and $\mathcal{L}$ are two $m$ dimensional subspace of $\mathbb{R}^n$. A projection technique onto the subspace $\mathcal{K}$ and orthogonal to $\mathcal{L}$ is a process which finds an approximate solution $\widetilde{\bm{X}}$ to Eq.~\ref{AXB} by imposing the conditions that $\widetilde{\bm{X}}$ belong to $\mathcal{K}$ and that the new residual vector be orthogonal to $\mathcal{L}$,
\begin{equation}\label{projection}
    Find\quad\widetilde{\bm{X}} \in \mathcal{K},\quad such\ that\ \bm{B} - \bm{A}\widetilde{\bm{X}} \bot \mathcal{L}.
\end{equation}
In a GMRES method, the $m$ dimensional subspace $\mathcal{K}$ is the $m$ dimension Krylov subspace formed as
\begin{equation}\label{mKrylov}
    \mathcal{K}_m(\bm{A}, v) = span\{\bm{r}_0, \bm{A}\bm{r}_0,\bm{A}^2\bm{r}_0,\ldots,\bm{A}^{m-1}\bm{r}_0\},
\end{equation}
where $\bm{r}_0 = \bm{B} - \bm{A}\bm{X}_0$ and $\bm{X}_0$ is the initial guess of solution. The subspace $\mathcal{L}$ is defined
as
\begin{equation}\label{GMRESLm}
    \mathcal{L}_m = \bm{A}\mathcal{K}_m.
\end{equation}
Since the obvious basis, $\bm{r}_0$. $\bm{A}\bm{r}_0$, \ldots, $\bm{A}^{m-1}\bm{r}_0$, of $\mathcal{K}_m$ is not very attractive from a numerical point of view, the kernel of GMRES method is to construct a group of orthogonal basis, $v_1$, \ldots, $v_m$, for the subspace $\mathcal{K}_m$. Let
\begin{equation}\label{GMRESVm}
	V_m = {v_1, v_2, \ldots, v_m},
\end{equation}
then the approximation can be expressed as $\widetilde{\bm{X}} = \bm{X}_0 + V_my_m$, where $y_m$ minimizes the function
$J(y)=\lVert\beta e_1-\overline{H}_m y\rVert_2$, i.e.,
\begin{equation}
    \widetilde{\bm{X}} = \bm{X}_0 + V_my_m,
\end{equation}
\begin{equation}
	J(y)=\lVert\beta e_1-\overline{H}_m y\rVert_2.
\end{equation}
Where $\beta = \lVert \bm{r}_0\rVert_2$, $v_1 = \bm{r}_0/\beta$, and $\overline{H}_m$ is the $(m+1)\times m$ Hessenberg matrix. Thus, the general procedure of the GMRES method for solving Eq.~\ref{AXB} can be summarized as follows
\begin{enumerate}
    \item Guessing an initial solution $\bm{X}_0$ for Eq.~\ref{AXB};
    \item Constructing a group of orthogonal basis,$v_1$,...,$v_m$, for the subspace $\mathcal{K}_m$, and the modified
        Gram-Schmidt method is always employed in this stage;
    \item Minimizing the function $J(y)=\lVert\beta e_1-\overline{H}_m y\rVert_2$, and finding $y_m$;
    \item Obtaining the approximate solution $\widetilde{\bm{X}} = \bm{X}_0 + V_my_m$;
    \item Checking whether $\widetilde{\bm{X}}$ can satisfy the Eq.~\ref{AXB}, if Eq.~\ref{AXB} is satisfied, the
        solution is obtained; if not, let $\bm{X}_0 = \widetilde{\bm{X}}$, then go to stage 1.
\end{enumerate}
The details of GMRES method can be referred to \cite{saad1986gmres}\cite{saad2003iterative}\cite{van2003iterative}.

\section{Numerical results and discussions}\label{Cases}
The gas-kinetic scheme proposed by Xu\cite{xu2001gas} is a unified methods which can be used for both incompressible and compressible flows. In our study, we develop a dual time-stepping strategy for gas-kinetic scheme, which is proved to be successful in the numerical methods based on Navier-Stokes equations. Three test cases are set up in this section, and they are used to demonstrate that the dual time-stepping method is not only useful for both incompressible and compressible flows, but also for laminar and turbulent flows.

The source code  based on our proposed algorithm is deployed on the Stanford University Unstructured (SU2) open-source platform\cite{palacios2013stanford}\cite{palacios2014stanford}. We appreciate the development team of SU2 for their great work.

\subsection{Case 1: Incompressible laminar flow around a circular cylinder}
The laminar flow past a single stationary circular cylinder, which has been studied using many  experimental and numerical methods \cite{park1998numerical}\cite{yuan2015immersed}\cite{tritton1959experiments}, is a benchmark of unsteady flows. In our paper, the aim of this test case is to validate the time-stepping strategy in the prediction of unsteady incompressible laminar flows.

In this case, the free-stream Mach number is $Ma_\infty = 0.1$, the Reynolds number are $Re_\infty = 60, 80, 100$, and the definition of Reynolds number is read as
\begin{equation}
	Re_\infty = \frac{\rho_\infty U_\infty d}{\mu_\infty},
\end{equation}
where, $d$ represents the diameter of the circular cylinder, $\rho_\infty$, $U_\infty$, and $\mu_\infty$ denote the density, velocity and the laminar viscosity of free stream flow respectively.

\begin{figure}[ht]
	\centering
	\includegraphics[width=0.4 \textwidth]{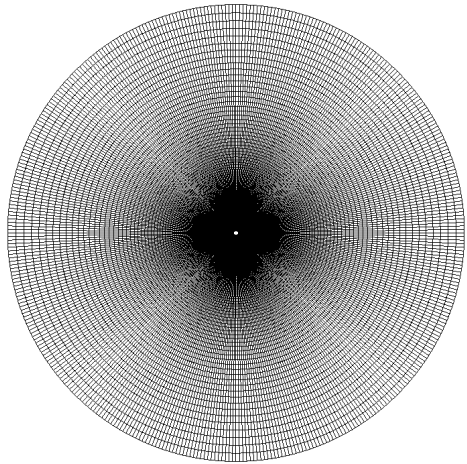}
    \caption{\label{figCylinderGrid} The computational domain of flow around a circular cylinder. The diameter of circular cylinder is $d=1$, and the outer diameter of the computational domain is $50d$.}
\end{figure}

The computational domain shown in Fig.~\ref{figCylinderGrid} is divided into an O-type grid, which has 400 points on the cylinder surface and 200 points on the radial direction. Characteristic information (Riemann invariants) based far-field boundary condition is applied on the outer of computational domain, and the no-slip and adiabatic wall condition is enforced on the surface of cylinder. The nearest distance of mesh points from the wall is $0.001$, and the y-plus is about $0.2$.

Table \ref{tableCircularCylinderCdCl} shows the comparison of drag coefficients and lift coefficients at different Reynolds numbers. $\overline{C_d}$ denotes the time-averaged total drag coefficient, $\widetilde{C_d}$ represents the fluctuations of drag coefficients away from $\overline{C_d}$, and $\widetilde{C_l}$ is the amplitude of the fluctuations of lift coefficients. The compared data come from other numerical methods and
experiments\cite{park1998numerical}\cite{tritton1959experiments}\cite{yuan2015immersed},
and the results demonstrate a good agreement with the referenced data.  Table \ref{tableCircularCylinderCdCl} shows that both the fluctuations of $C_d$ and $C_l$ are evident, and it is clear that the fluctuations of lift coefficient are much bigger than drag coefficient. As $Re_\infty$ increases, the amplitude of fluctuations of total drag coefficient and lift coefficient increase, but the total drag coefficient decreases.

\begin{table}[!htp]
	\centering
	\caption{\label{tableCircularCylinderCdCl} Drag and lift coefficients of a circular cylinder.}
	\begin{threeparttable}
        \begin{tabular}{p{20pt} p{40pt}<{\centering} p{40pt}<{\centering} p{40pt}<{\centering} p{60pt}
            <{\centering} p{60pt}<{\centering} p{40pt}
            <{\centering} p{40pt}<{\centering} p{40pt}<{\centering}}
			\hline
			\hline
            & \multicolumn{3}{c}{Park et al.\cite{park1998numerical}}& Yuan\cite{yuan2015immersed} & Tritton\cite{tritton1959experiments} &\multicolumn{3}{c}{Present} \\
            \cline{2-4}   \cline{7-9} \\
            $Re_\infty$ & $\overline{C_d}$ & $\widetilde{C_d}$ & $\widetilde{C_l}$& $\overline{C_d}$ & $\overline{C_d}$
            & $\overline{C_d}$ & $\widetilde{C_d}$ & $\widetilde{C_l}$ \\

            $60$ & $1.39$ & $0.0014$ & $0.1344$&$1.419$ & $1.398$& $1.39$ & $0.0010$ & $0.1248$ \\

            $80$ & $1.35$ & $0.0049$ & $0.2452$ &$1.376$& $1.316$& $1.35$ & $0.0042$  & $0.2355$\\

            $100$ & $1.33$ & $0.0091$ & $0.3321$ &$1.352$& $1.271$&$1.33$ & $0.0093$ & $0.3220$ \\
            \lasthline
		\end{tabular}
	\end{threeparttable}
\end{table}

In dimensional analysis, the Strouhal number $St$ is a non-dimensional number which describes the vortex shedding frequency  of unsteady flows, and it is defined in our paper as
\begin{equation}
	St = \frac{fd}{U_\infty}.
\end{equation}
Where, $f$ denotes the vortex shedding frequency. C. Williamson gives an approximative formula for Strouhal number versus Reynolds number for circular cylinders \cite{williamson1989oblique}, which can be expressed as
\begin{equation}\label{St}
	St = -3.3265/Re + 0.1816 + 1.6 \times 10^{-4}Re.
\end{equation}
The Strouhal numbers investigated in our paper are compared with data from other researchers. Table \ref{tableCircularCylinderSt}
gives the details of Strouhal number in our study, and the results shows a good accordance with compared data.
\begin{table}[!htp]
	\centering
	\caption{\label{tableCircularCylinderSt} The comparison of Strouhal number of a circular cylinder.}
	\begin{threeparttable}
		\begin{tabular}{p{80pt} p{25pt}<{\centering} p{120pt}<{\centering} p{80pt}<{\centering}}
			\hline
			\hline
			$Re_\infty$ & Present & Williamson\cite{williamson1989oblique} & Silva\cite{silva2003numerical} \\
			\hline
			$60$ & $0.1329$ & $0.1358$ & - \\

			$80$ & $0.1509$ & $0.1528$ & $0.1495$ \\

			$100$ & $0.1621$ & $0.1643$ & $0.1615$ \\		
			\lasthline
		\end{tabular}
	\end{threeparttable}
\end{table}

\begin{figure}[!htp]
	\centering
	\subfigure[]{
		\includegraphics[width=0.45 \textwidth]{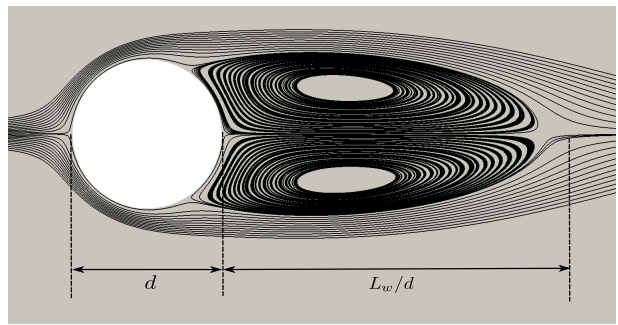}
		\label{figCircularCylinderLwha}
	} \\
	\subfigure[]{
		\includegraphics[width=0.45 \textwidth]{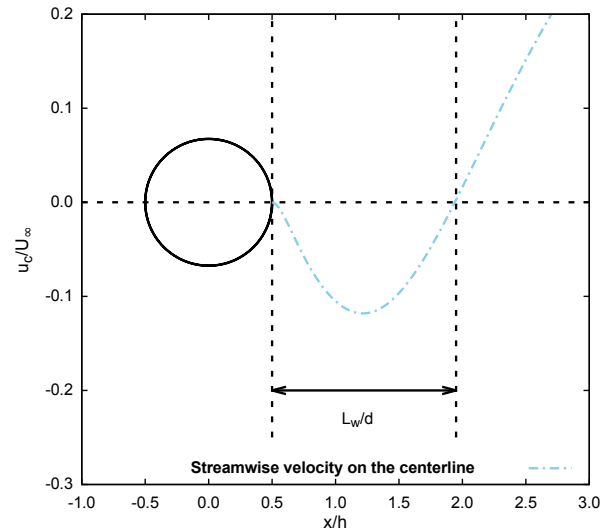}
		\label{figCircularCylinderLwhb}
	}
	\caption{The definition of  length of recirculation bubble ($L_w$).}
	\label{figCircularCylinderLwh}
\end{figure}

\begin{figure}[!htp]
	\centering
	\subfigure[]{
		\includegraphics[width=0.4\textwidth]{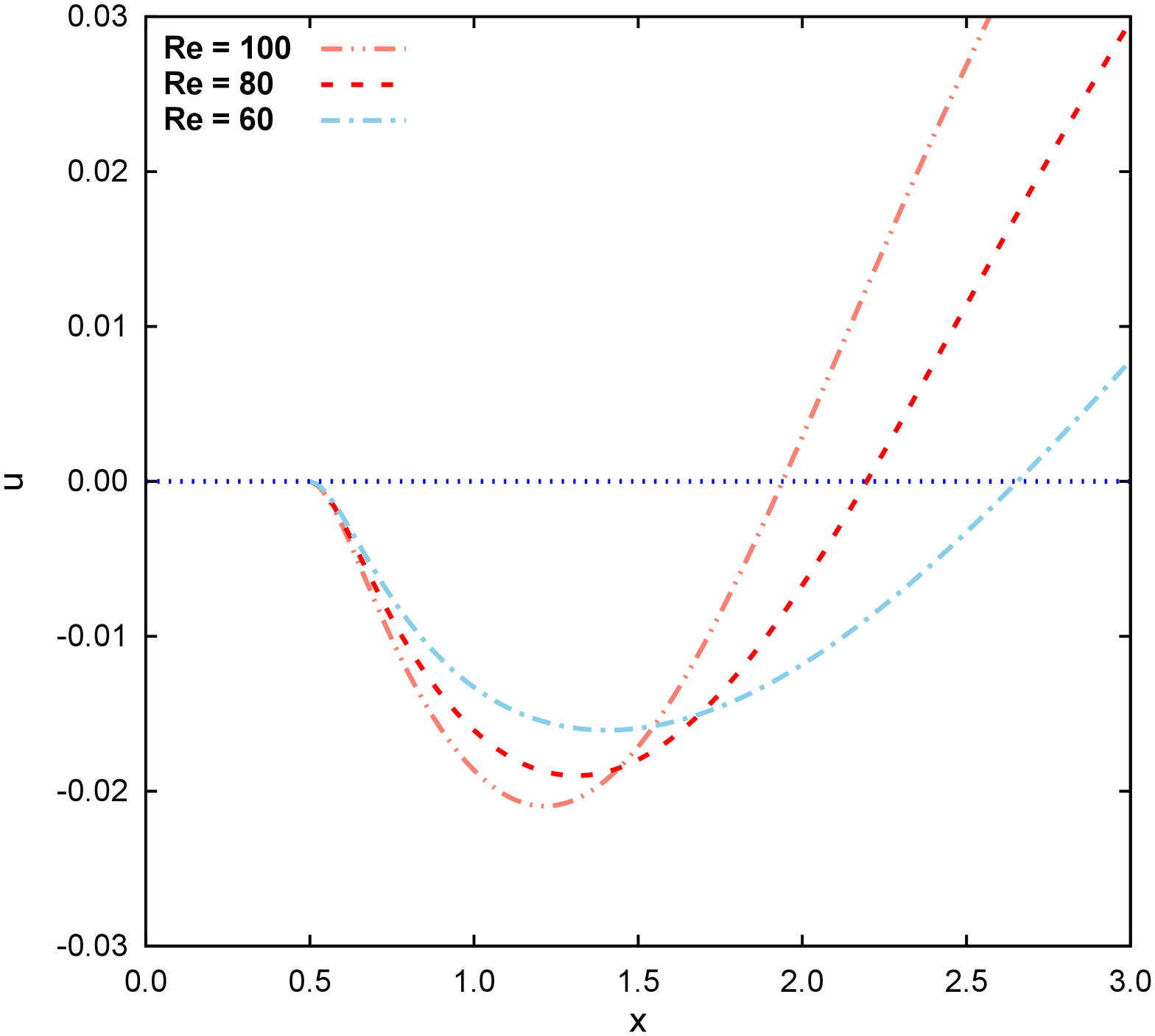}
		\label{figCircularCylinderU}
	}
	\subfigure[]{
		\includegraphics[width=0.4\textwidth]{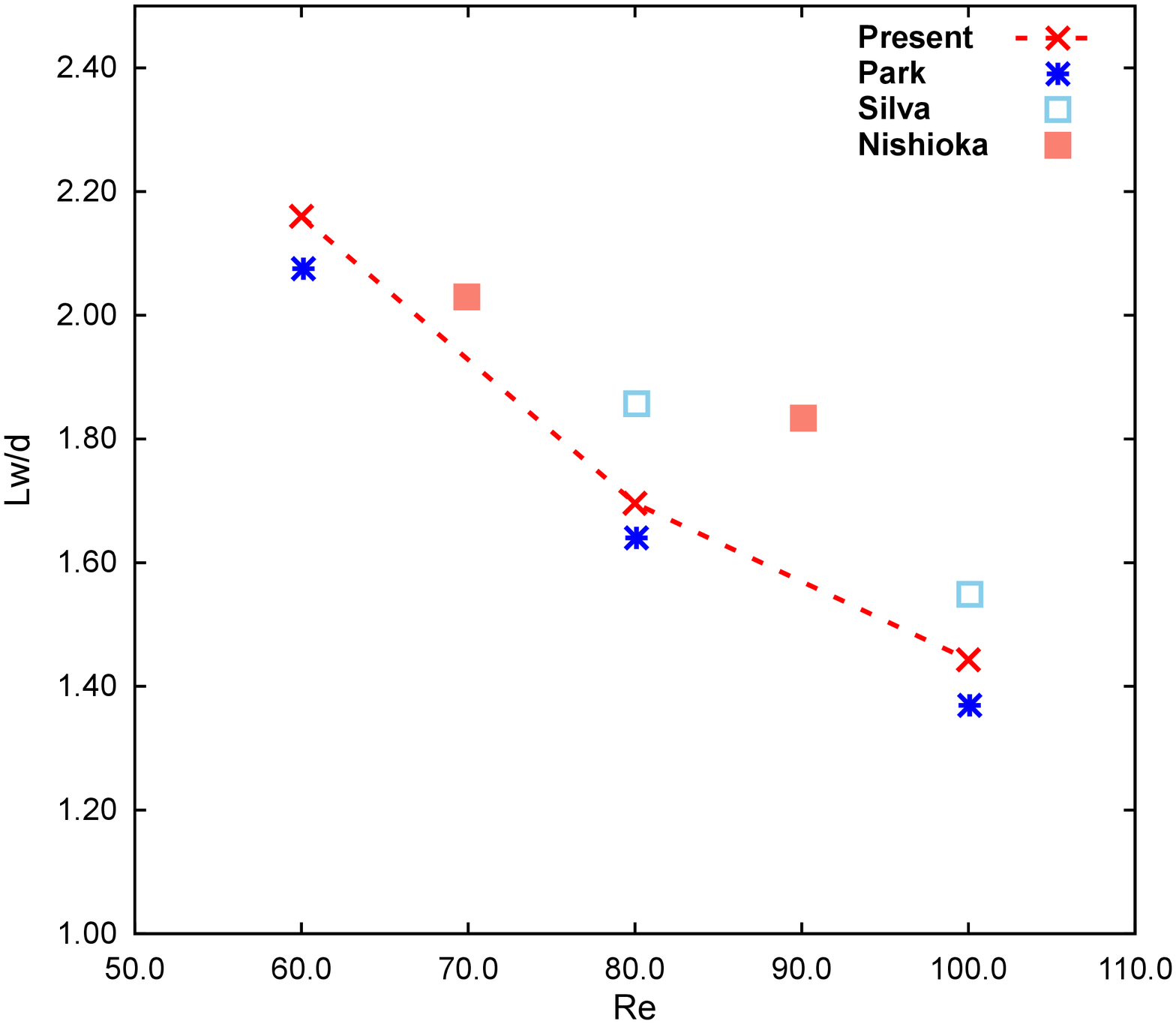}
		\label{figCircularCylinderLw}
	}
	\caption{\label{figCircularCylinderLwU} The mean velocity at the central line and length of recirculation bubble vs. Reynolds numbers. (a) the mean velocity at the central line. (b) the comparison of length of recirculation bubble.}
\end{figure}

The length of the recirculation bubble $L_w$ is defined as the distance between two stagnation points downstream of the cylinder. For an unsteady flow, the determination of $L_w$ defined in  Fig.~\ref{figCircularCylinderLwh} is based on the mean flow field in a long time interval.  In our study, we use the horizontal velocity on the line $y = 0$ to calculate the length of recirculation bubble, and Fig.~\ref{figCircularCylinderU} plots the mean horizontal velocity at different Reynolds numbers. Fig.~\ref{figCircularCylinderLw} shows the comparison of $L_w$ with data by other numerical methods and experiments\cite{park1998numerical}\cite{silva2003numerical}\cite{nishioka1978mechanism}.

\begin{figure}[!htp]
	\centering
	\includegraphics[width=0.4 \textwidth]{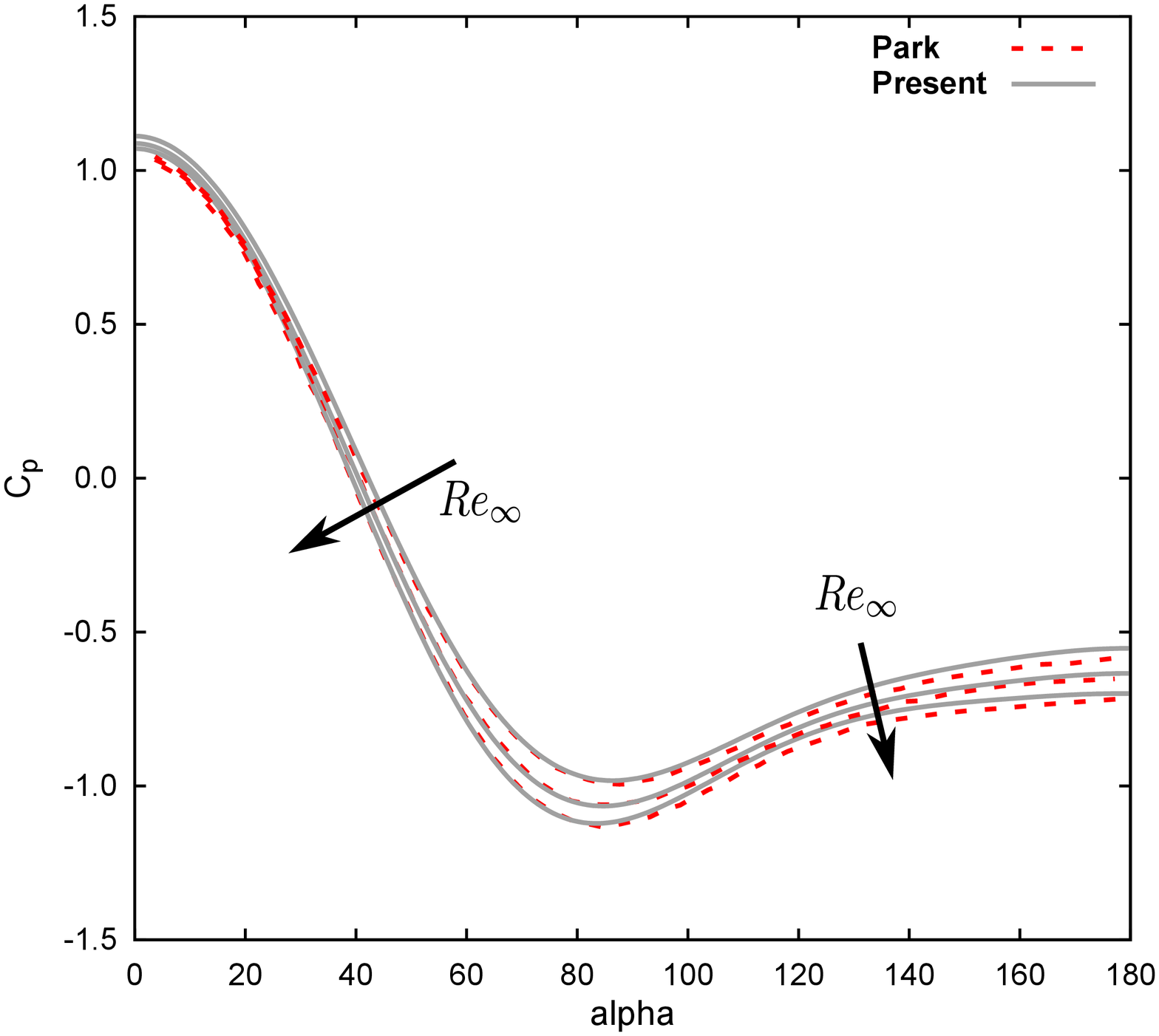}
    \caption{\label{figCircularCylinderCp} Pressure coefficient on the cylinder surface at Reynolds numbers $Re_\infty= 60, 80, 100$.}
\end{figure}

The pressure coefficients of mean flow field at different Reynolds numbers on the cylinder surface are shown in Fig.~\ref{figCircularCylinderCp}, where $\theta = 0^{\circ}$ and $\theta = 180^{\circ}$ correspond to  the stagnation and base points respectively. The plots demonstrate a good accordance with the compared data by Park\cite{park1998numerical}.

\begin{figure}[!htp]
	\centering
	\subfigure[t=0]{
		\includegraphics[width=0.4\textwidth]{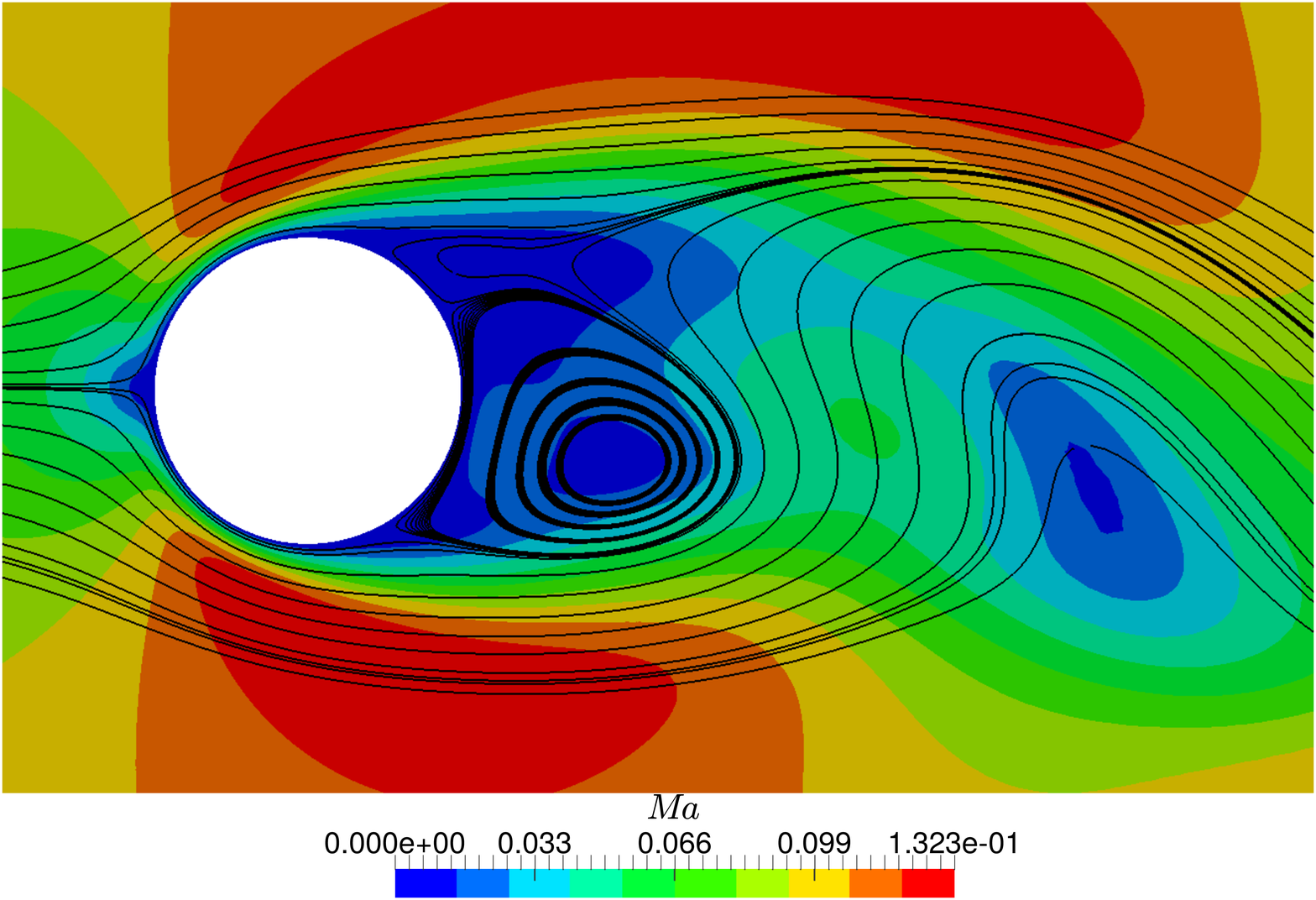}
		\label{figCircularCylinderT0}
	}
	\subfigure[t=$\frac{T}{4}$]{
		\includegraphics[width=0.4\textwidth]{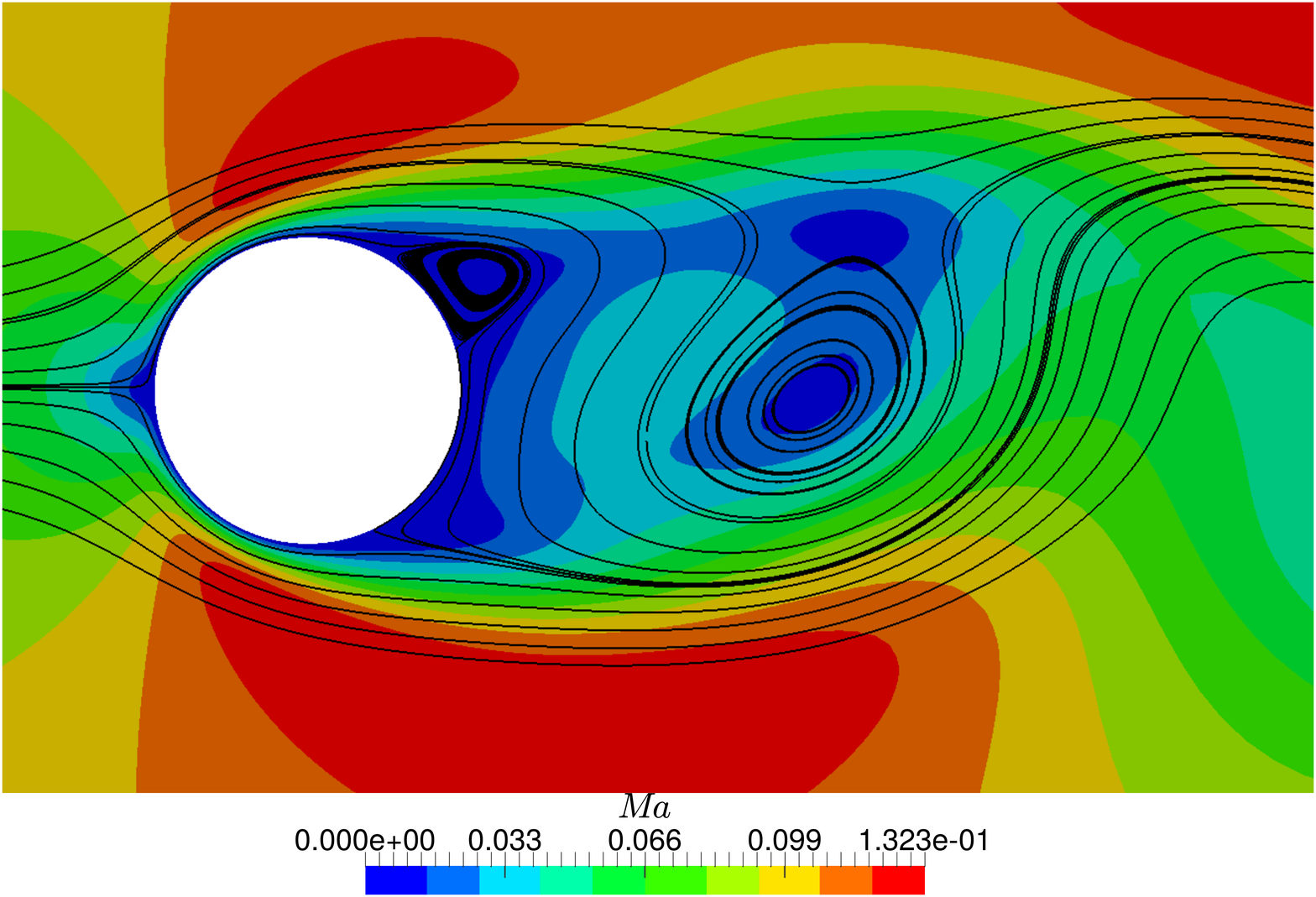}
		\label{figCircularCylinderT1}
	}
	\subfigure[t=$\frac{T}{2}$]{
		\includegraphics[width=0.4\textwidth]{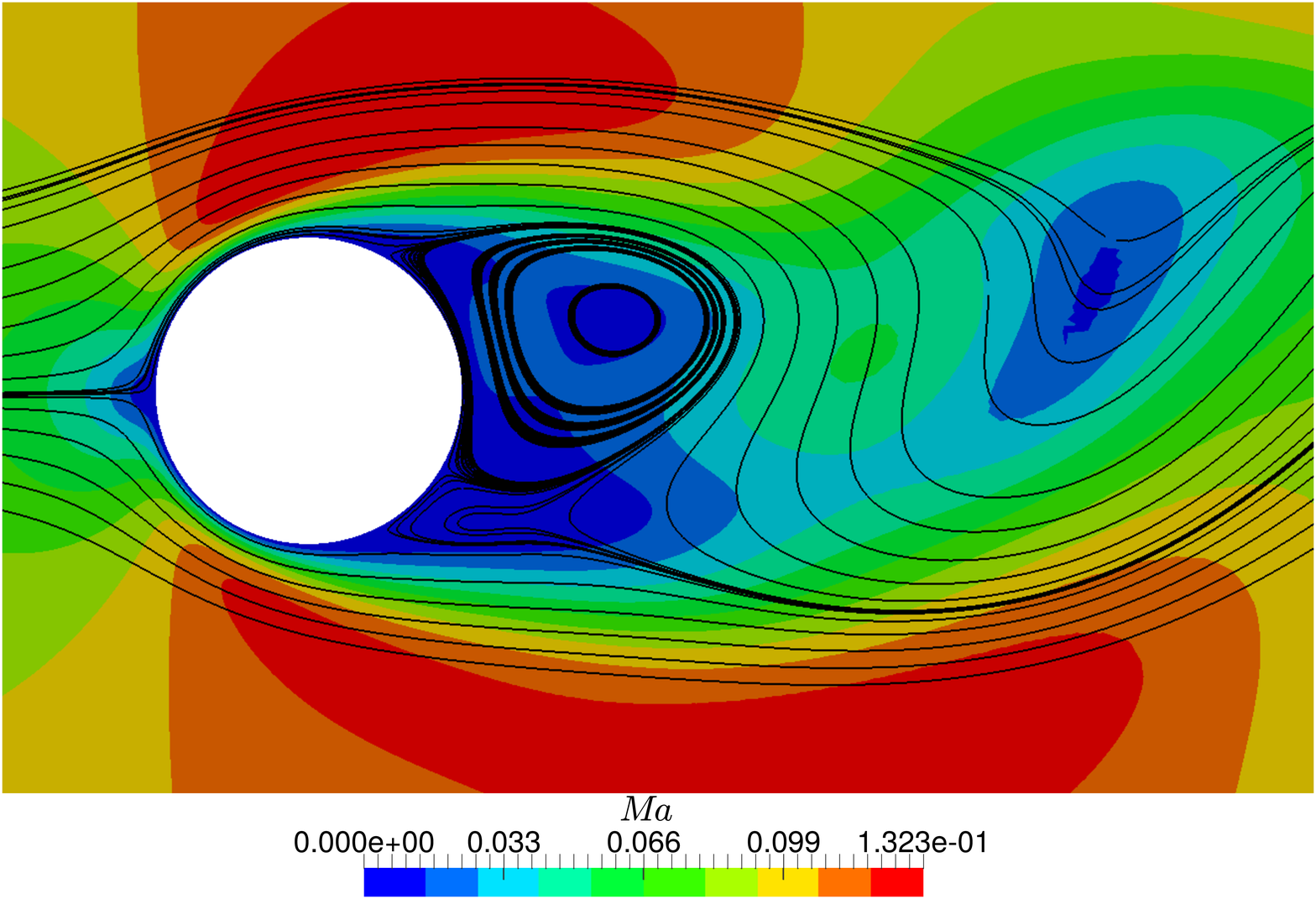}
		\label{figCircularCylinderT2}
	}
	\subfigure[t=$\frac{3T}{4}$]{
		\includegraphics[width=0.4\textwidth]{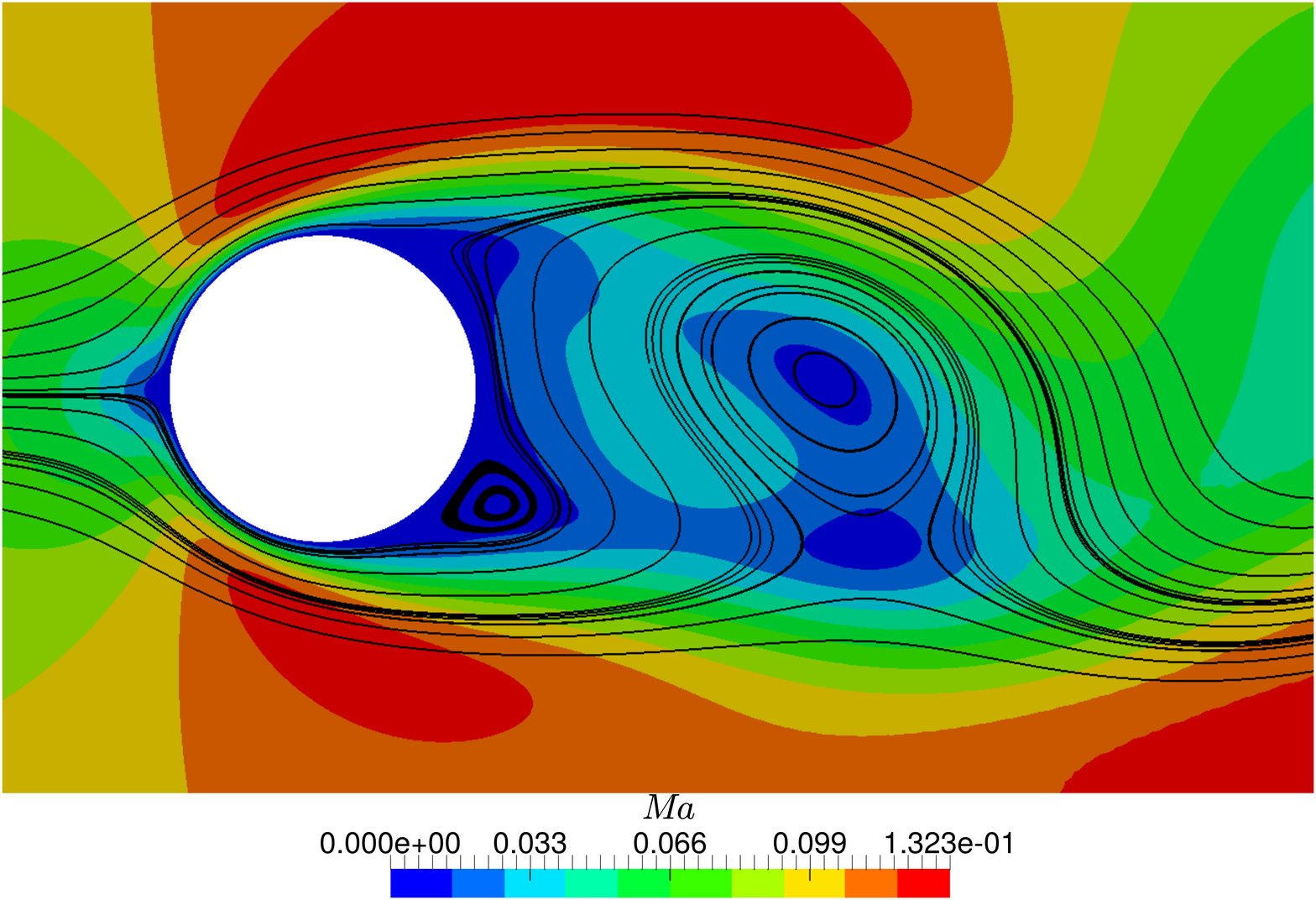}
		\label{figCircularCylinderT3}
	}
	\caption{Time history of the streamlines past a circular cylinder. T represents the period.}
	\label{figCircularCylinderT}
\end{figure}

A qualitative picture of flow streamlines, $Re_\infty = 100$, laid over a Mach number contour plots is presented in Fig.~\ref{figCircularCylinderT}. As expected, the periodic vortex shedding can be seen clearly in the wake of circular cylinder. It is obvious that the vortices are shed alternative from each side of the circular cylinder, and then converted down stream in the wake of the cylinder.

\begin{table}[!htp]
	\centering
	\caption{\label{tableCircularCylinderWorks} The comparison of computational works between explicit scheme and dual
    time-stepping strategy ($Re_\infty=100$).}
	\begin{threeparttable}
        \begin{tabular}{p{90pt} p{40pt}<{\centering} p{80pt}<{\centering} p{70pt}<{\centering}}
			\hline
			\hline
            Scheme & $\Delta t$ & Inner iteration & Pseudo steady resudial\\
			\hline
            Explicit & $0.0015$ & $-$ & $-$ \\

            Dual time-stepping & $0.1$ & $10$ & $< 10^{-7}$ \\
			\lasthline
		\end{tabular}
	\end{threeparttable}
\end{table}

Table \ref{tableCircularCylinderWorks} shows the time step of explicit scheme and dual time-stepping strategy
respectively ($Re_\infty = 100$). The time step of explicit scheme is determined by Eq.~\ref{exDeltaT} and the time step of dual time-stepping method is the physical time step. To predict the flow field at time $t = 1500$, the explicit scheme needs $1\times 10^6$ steps, and the dual time-stepping method needs only $1.5\times 10^5$ steps. It is obvious that the dual time-stepping strategy of gas-kinetic scheme can save a lot of computational works in the approach of unsteady incompressible flows, and the residual of pseudo steady solution is sufficient to guarantee the accuracy of the dual time-stepping method for flow simulation of unsteady flows.

\subsection{Case 2: Incompressible turbulent flow around a square cylinder}
The incompressible turbulent flow around a square cylinder is investigated in this section. The case is studied by many numerical methods\cite{franke1993calculation}\cite{iaccarino2003reynolds}\cite{rodi1997comparison}\cite{bosch1998simulation} and experiments\cite{durao1988measurements}\cite{lyn1995laser}\cite{vickery1966fluctuating}\cite{lee1975effect}. In our paper, we explore it using gas-kinetic scheme coupled with Menter's Shear Stress Transport (SST) turbulence model\cite{menter1994two}, and a gas-kinetic scheme coupled with SST turbulence model has been introduced in the Ref. \cite{li2016gas}.
	
The aim of this test case is to examine the behavior of dual time-stepping method on the incompressible turbulent flow.	At the beginning of the simulation, an incompressible free stream flow with $Ma_\infty = 0.15$ and $Re_\infty = 22,000$ is initiated in the computational domain.  With the time evolution, the unsteady phenomena appear inside the flow field. The Reynolds number $Re_\infty$ is defined as
\begin{equation}
Re_\infty = \frac{\rho_\infty U_\infty h}{\mu_\infty},
\end{equation}
and $h$ represents the side length of a square cylinder. For a turbulent flow, a small, $2\%$, turbulence intensity is imposed in the inlet, and the ration of eddy viscosity and laminar viscosity equals $0.1$ in the far field.

\begin{figure}[!htp]
	\centering
	\includegraphics[width=0.6 \textwidth]{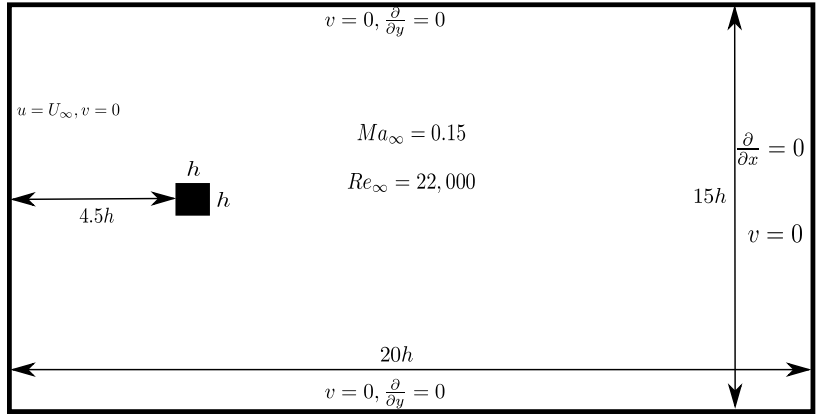}
	\caption{\label{figsqCylinderDomain} Computational domain and boundary conditions for the simulation of incompressible flow around a square cylinder.}
\end{figure}

The computational domain is a $20h \times 15h$ rectangle. The square cylinder is located at $(5h,7.5h)$. The boundary conditions used in the approach are adopted from the study of Franke\cite{franke1993calculation}.  Fig.~\ref{figsqCylinderDomain} shows the details of computational domain and boundary conditions for the flow simulation.

\begin{figure}[!htp]
	\centering
    \subfigure[Full domain.]{
		\includegraphics[width=0.4 \textwidth]{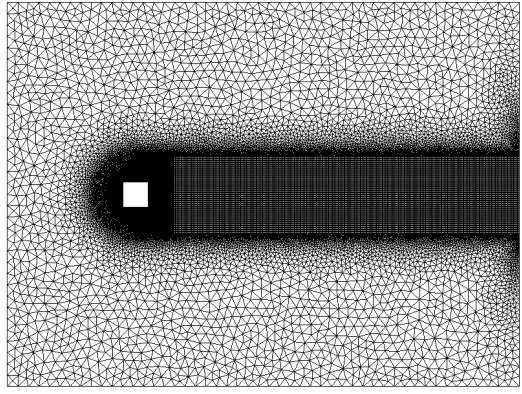}
		\label{figSqCylinderGrid1}
	} \\
	\subfigure[View of the grids near the cylinder surface.]{
		\includegraphics[width=0.4 \textwidth]{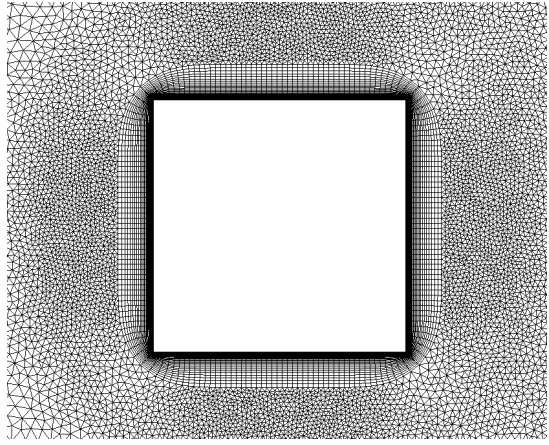}
		\label{figSqCylinderGrid2}
	}
	\caption{\label{figSqCylinderGrid} Grids for the simulation of incompressible flow around a square cylinder.}
\end{figure}

Fig.~\ref{figSqCylinderGrid} shows the hybrid grids used for the prediction of incompressible flow around a square cylinder. The grid is made up of rectangles and triangles, and the total number of cells in the domain is $70652$. The rectangular part distributed around the cylinder is used to guarantee the simulation accuracy inside viscous boundary layer, and the rectangular part in the wake of cylinder is used to obtain the accurate vortex frequency. The nearest distance from cylinder wall is $0.0008$, and the y-plus is $y^+ \approx 1.25$.
\begin{figure}[!htp]
	\centering
	\subfigure[]{
		\includegraphics[width=0.45 \textwidth]{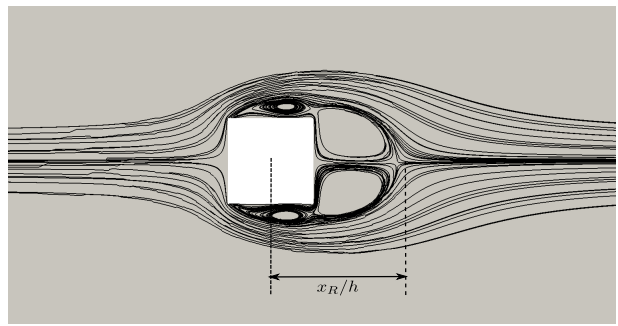}
		\label{figsqCylinderXR0}
	} \\
	\subfigure[]{
		\includegraphics[width=0.45 \textwidth]{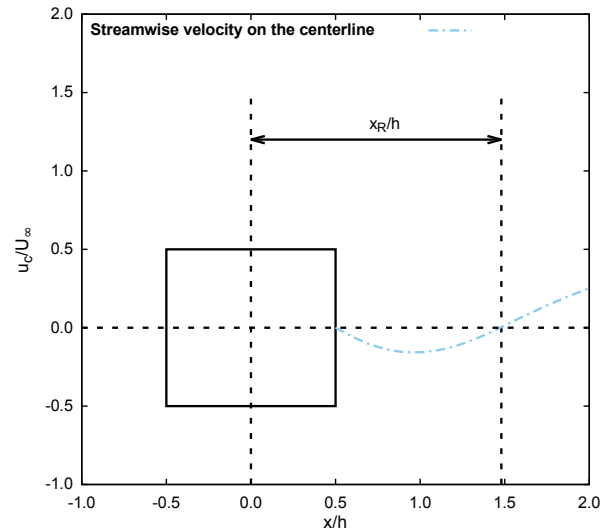}
		\label{figsqCylinderXR1}
	}
	\caption{The definition of the length of recirculation region ($x_R$).}
	\label{figsqCylinderXR}
\end{figure}
\begin{figure}[!htp]
	\centering
	\includegraphics[width=0.4 \textwidth]{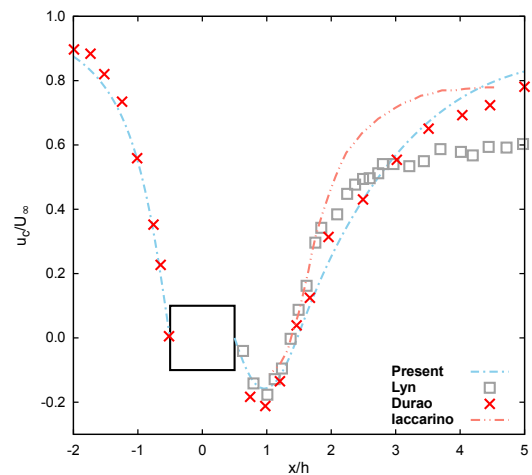}
	\caption{\label{figsqCylinderuc} Streamwise velocity profiles in the wake centerline of a square cylinder.}
\end{figure}

\begin{figure}[!htp]
	\centering
	\subfigure[$x=0.875$]{
		\includegraphics[width=0.4\textwidth]{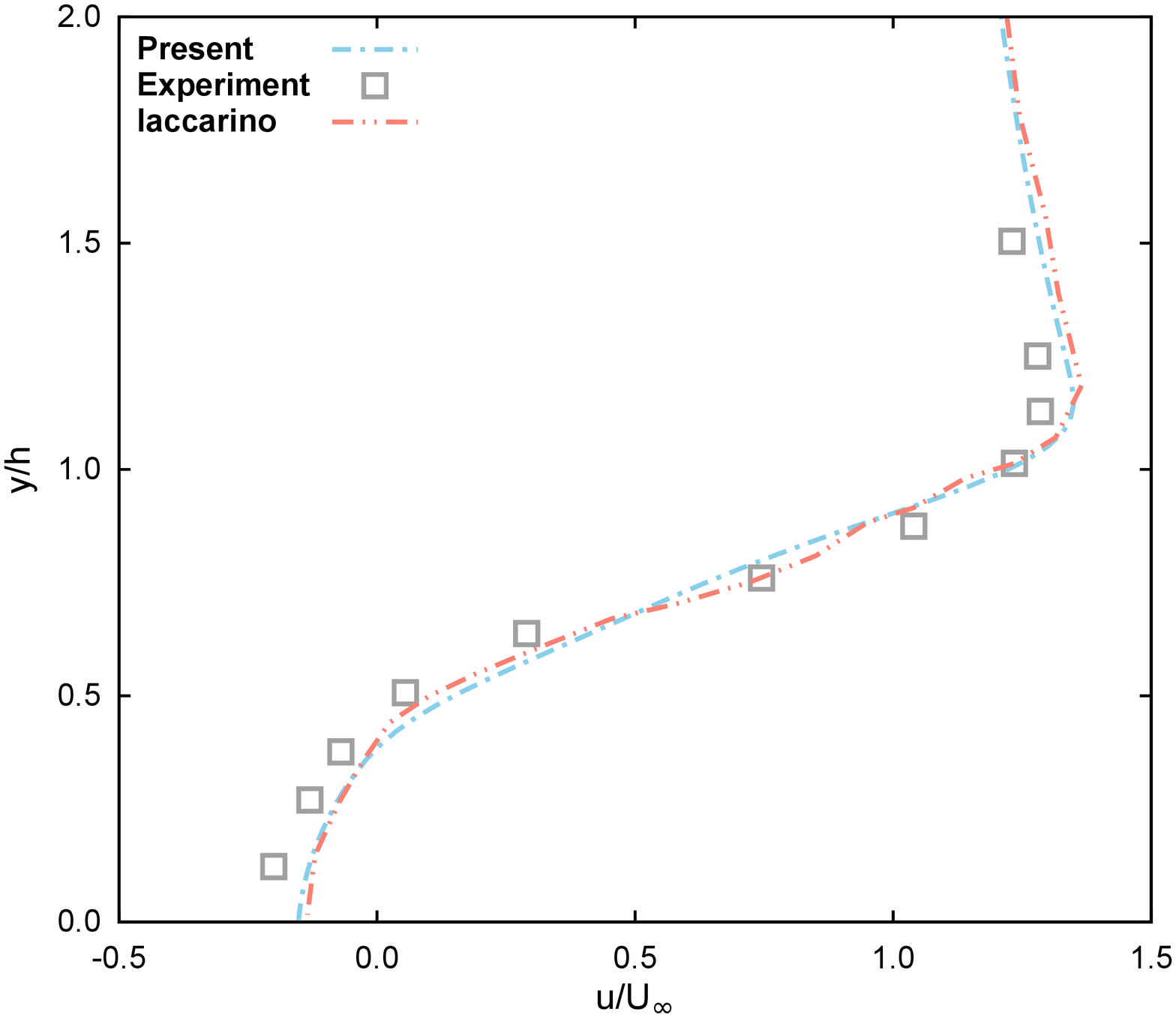}
		\label{figsqCylinderu4a}
	}
	\subfigure[$x=1.125$]{
		\includegraphics[width=0.4\textwidth]{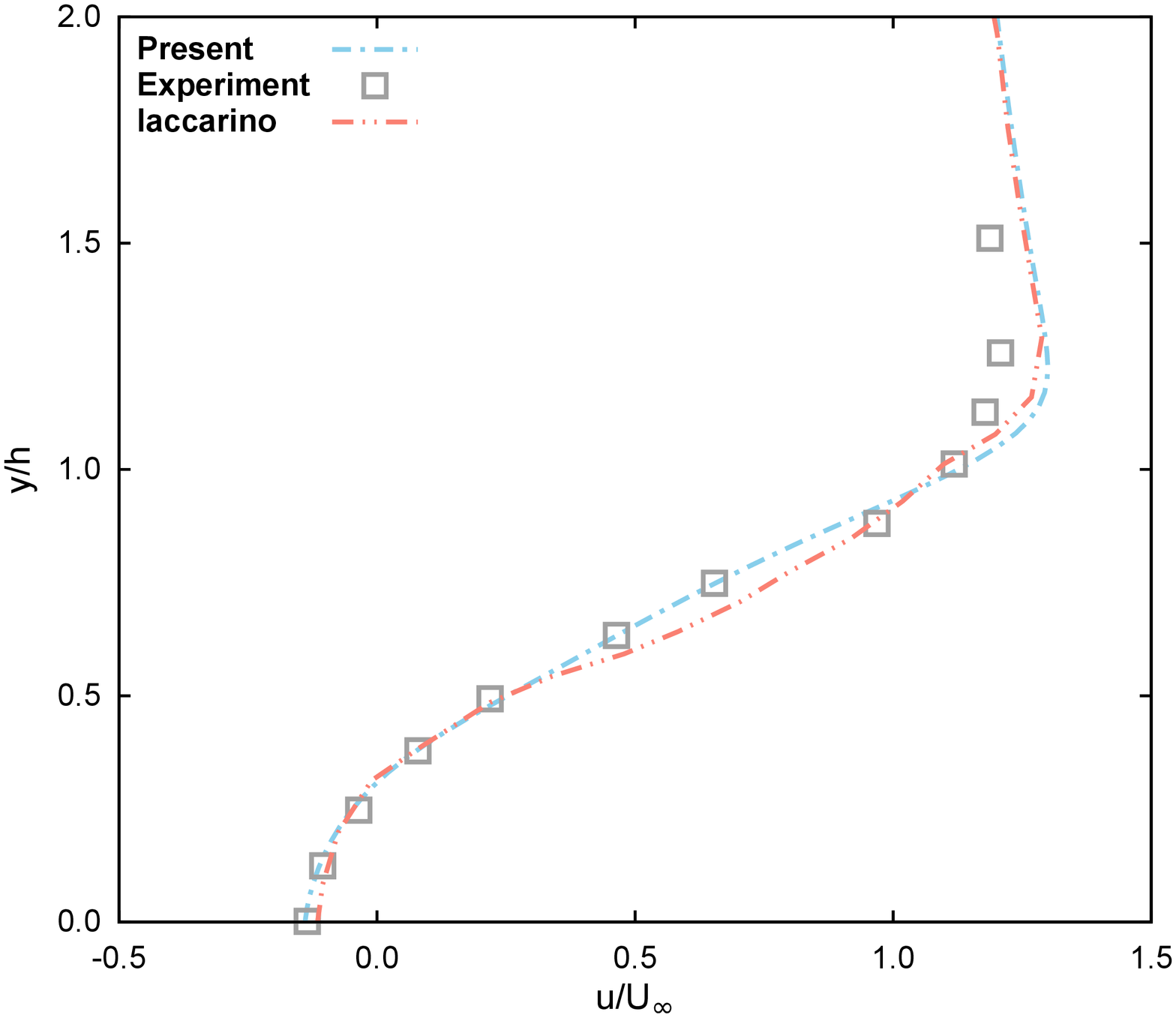}
		\label{figsqCylinderu4b}
	}
	\subfigure[$x=1.875$]{
		\includegraphics[width=0.4\textwidth]{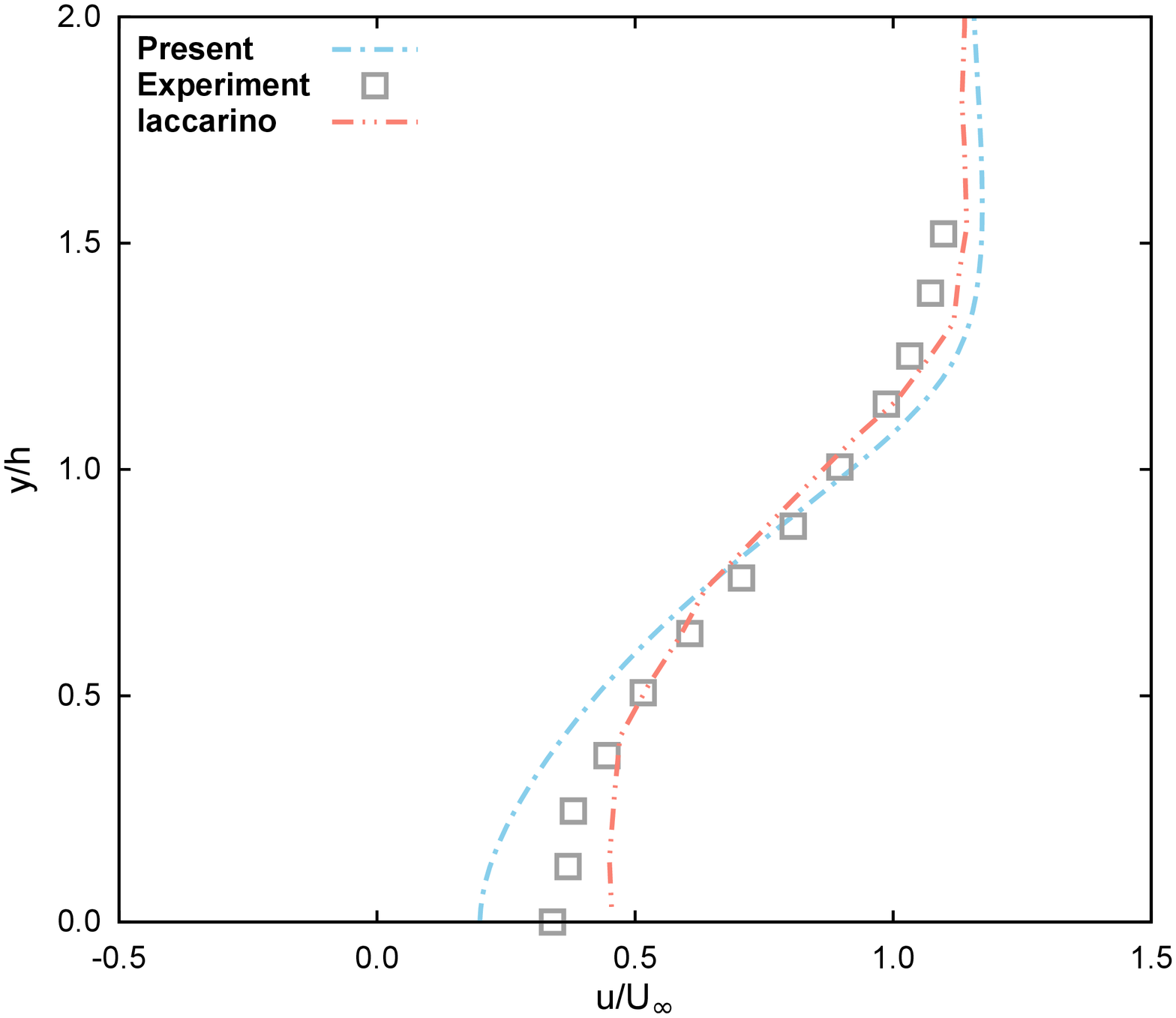}
		\label{figsqCylinderu4c}
	}
	\subfigure[$x=3.5$]{
		\includegraphics[width=0.4\textwidth]{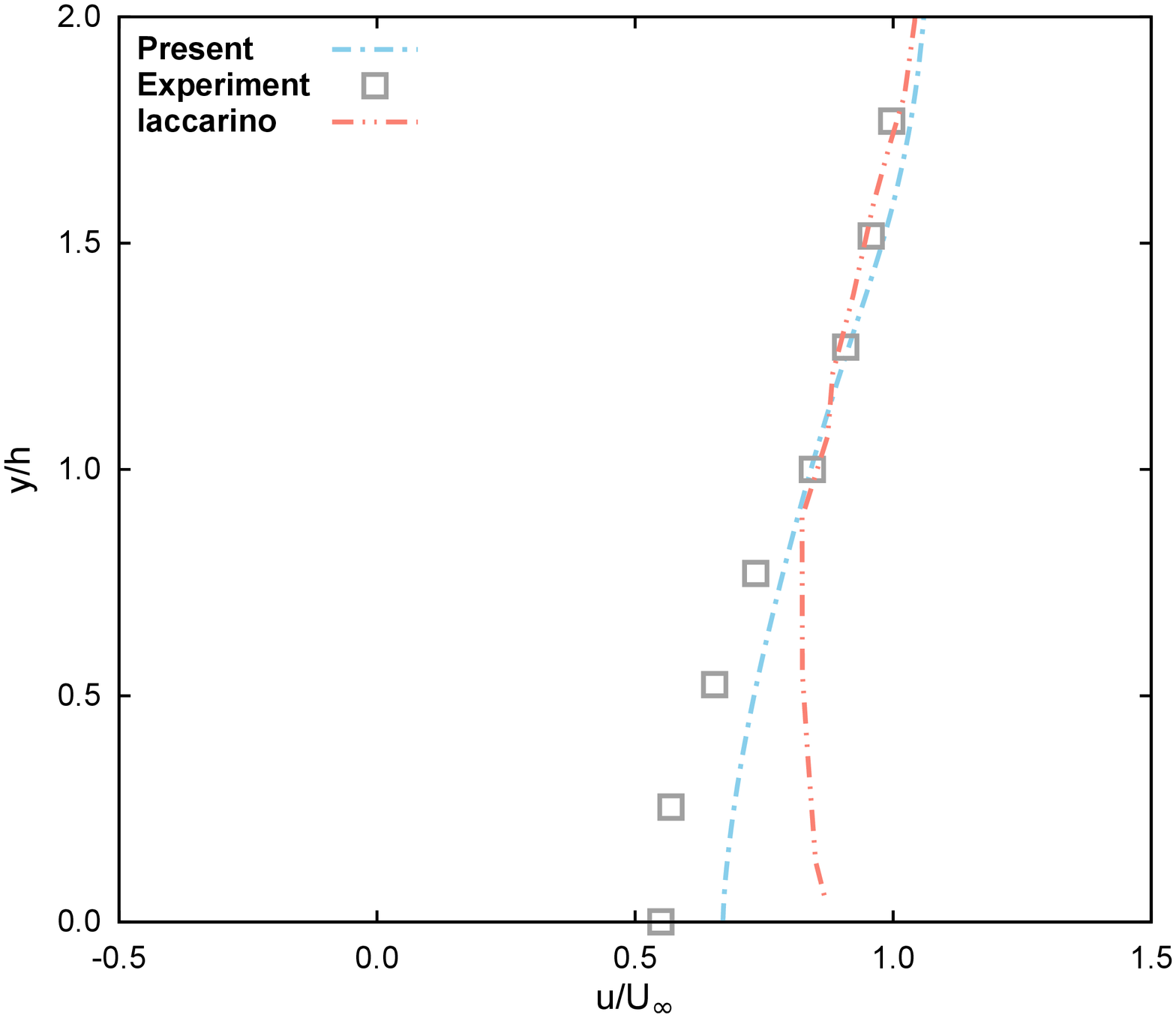}
		\label{figsqCylinderu4d}
	}
	\caption{Streamwise velocity profiles at four positions in the wake of a square cylinder.}
	\label{figsqCylinderu4}
\end{figure}
\begin{figure}[!htp]
	\centering
	\subfigure[$t=0$]{
		\includegraphics[width=0.4\textwidth]{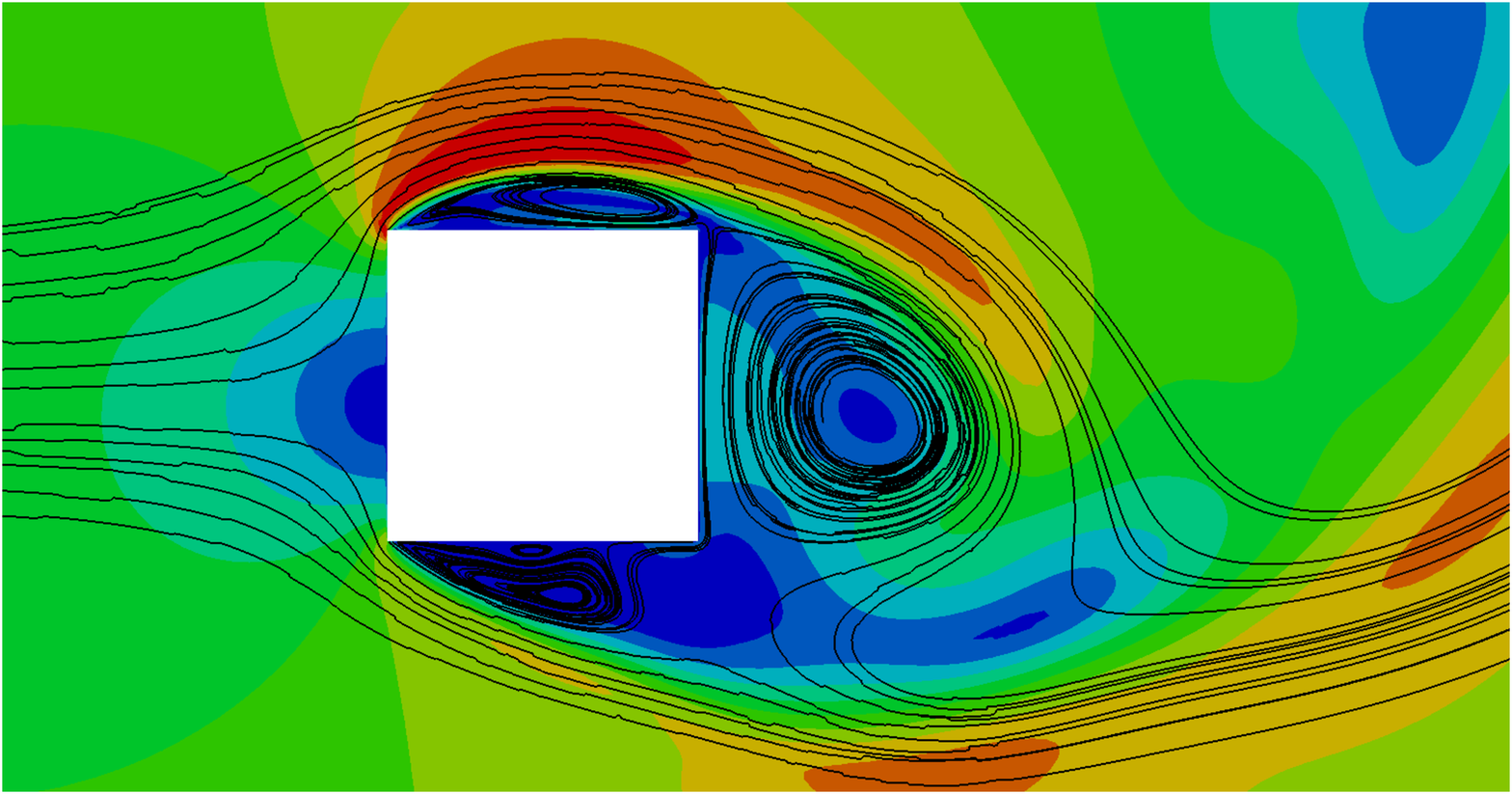}
		\label{figsqCylindeStreamline4a}
	}
	\subfigure[$t=\frac{T}{4}$]{
		\includegraphics[width=0.4\textwidth]{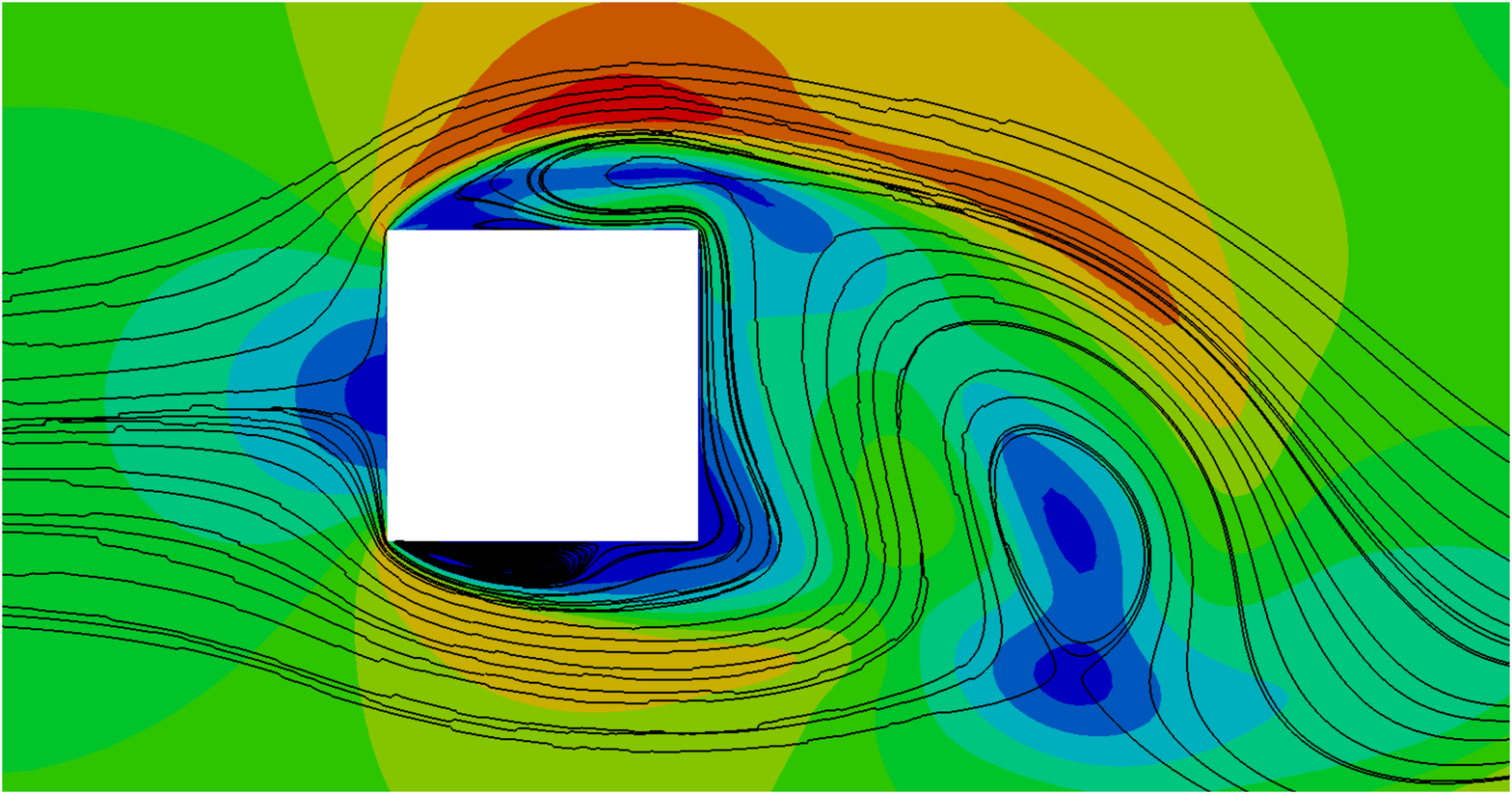}
		\label{figsqCylindeStreamline4b}
	}
	\subfigure[$t=\frac{T}{2}$]{
		\includegraphics[width=0.4\textwidth]{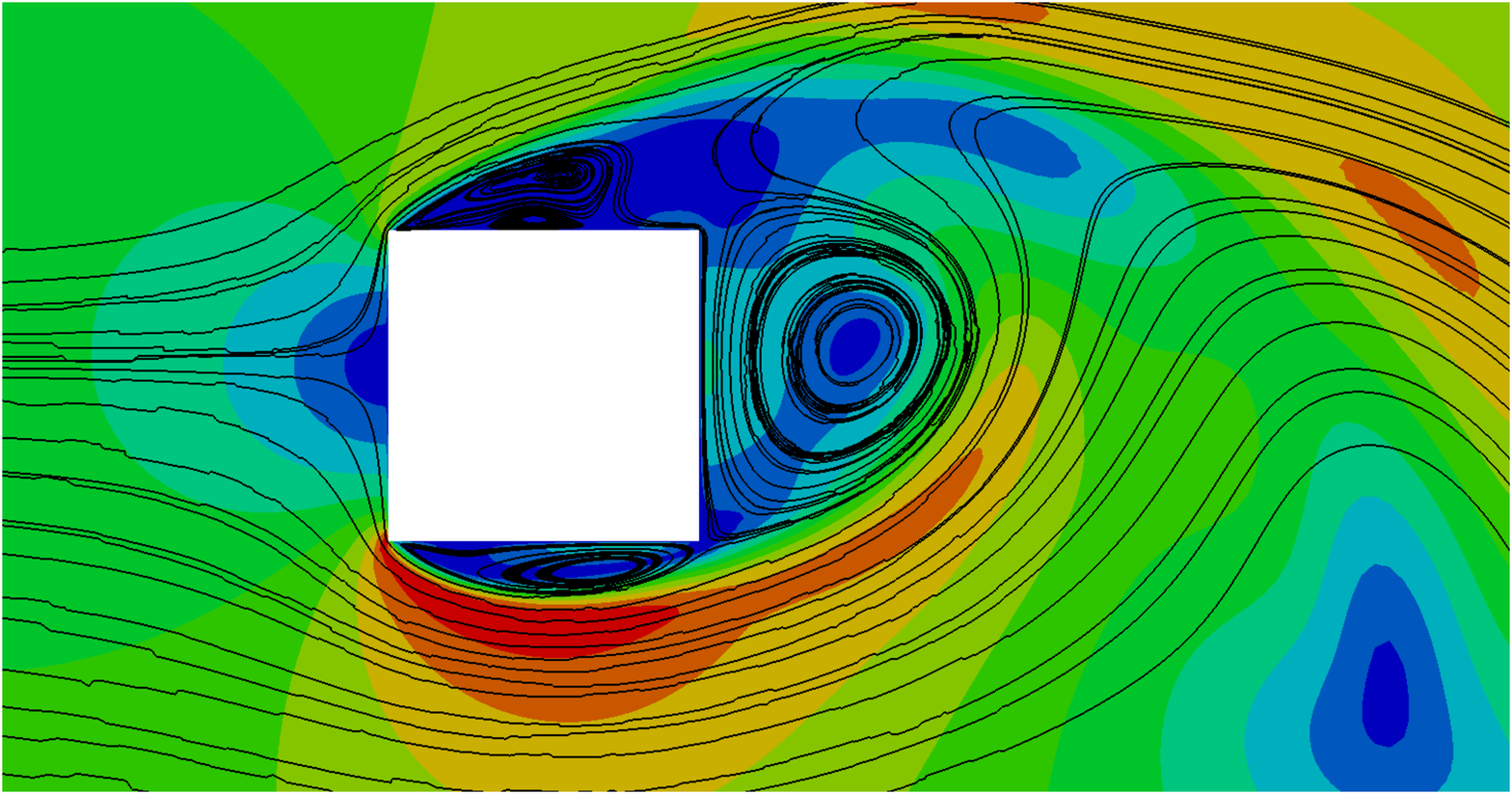}
		\label{figsqCylindeStreamline4c}
	}
	\subfigure[$t=\frac{3T}{4}$]{
		\includegraphics[width=0.4\textwidth]{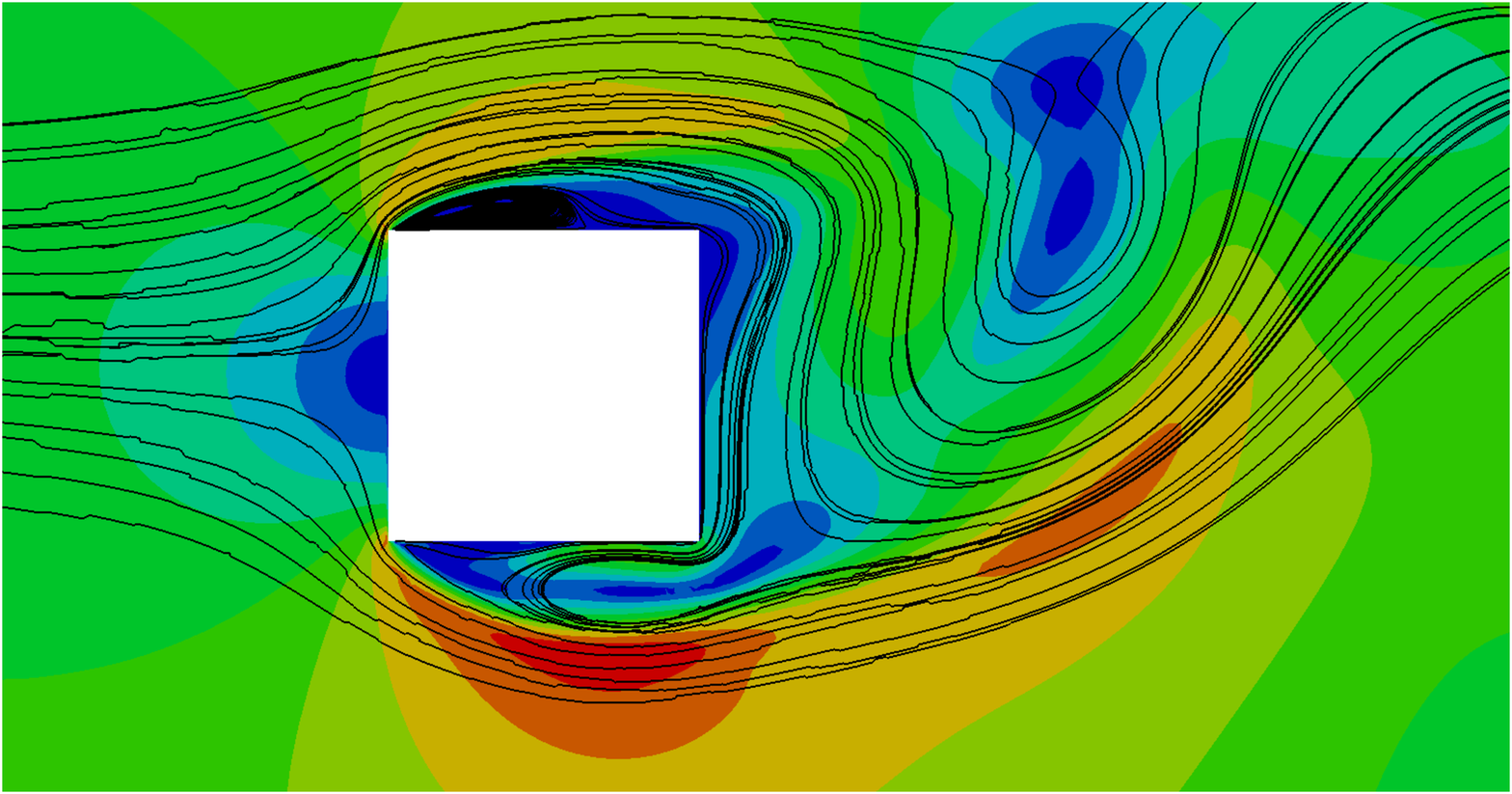}
		\label{figsqCylindeStreamline4d}
	}
	\caption{Time history of the streamlines in the wake of a square cylinder. T represents the period.}
	\label{figsqCylindeStreamline4}
\end{figure}
The incompressible turbulent flow ($Re_\infty = 22,000$) around a square cylinder which is investigated in our paper presents coherent vortex shedding with a periodically oscillating wake. A summary of data from present simulation, several numerical methods and experiments, are reported in Table \ref{tablesqCylinder}. $\overline{C_d}$ denotes the time averaged drag coefficient, and the Strouhal number $St$ is defined as
\begin{equation}\label{sqSt}
	St = \frac{fh}{U_\infty},
\end{equation}
where $f$ is the frequency of vortex shedding. $\widetilde{C_d}$ and $\widetilde{C_l}$ are the root mean square of drag and lift coefficients respectively.  The vortex shedding frequency represented by Strouhal number $St$ is in a good agreement with experimental and computational results found in the literature. One of important features that has to be analyzed is the length of recirculation region just downstream of a square cylinder. The recirculation region, which is formed due to the separation, is characterized by $x_R$, and the definition of $x_R$ is shown in Fig.~\ref{figsqCylinderXR}. To determine the value of $x_R$, mean flow field must be obtained in a long time interval. The value of $x_R$ in our study is in a good accordance with the data from experiments and other numerical methods. The surface loads are also of great importance.  It can be seen in Table \ref{tablesqCylinder} that the time-averaged drag coefficient in our simulation is acceptable compared with the other data.  $\widetilde{C_d}$ and $\widetilde{C_l}$ represent the fluctuations of drag and lift coefficient respectively, and both of them are in good accordance with the
compared data.

\begin{table}[!htp]
	\centering
	\caption{\label{tablesqCylinder} Results of a square cylinder.}
	\begin{threeparttable}
		\begin{tabular}{p{50pt}<{\centering} p{70pt}<{\centering} p{20pt}<{\centering} p{20pt}<{\centering} p{45pt}<{\centering} p{45pt}<{\centering} p{10pt}<{\centering} p{40pt}}
			\hline
			\hline
			Contribution & Model & $x_R/h$ & $\overline{C_d}$ & $\widetilde{C_d}$ & $\widetilde{C_l}$& $St$ \\
			\hline
			Present & SST model & $1.45$ & $2.02$ & $0.234$ & $1.134$ & $0.124$\\

			Lyn\cite{lyn1995laser} & Experiments & $1.38$ & $2.1$ & $-$ & $-$ & $0.132$\\

			Lee\cite{lee1975effect} & Experiments & $-$ & $2.05$ & $0.16-0.23$ & $-$ & $-$\\

			Vickery\cite{vickery1966fluctuating} & Experiments & $-$ & $2.05$ & $0.1-0.2$ & $0.68-1.32$ & $-$\\

			Iaccarino\cite{iaccarino2003reynolds} & Unsteady & $1.45$ & $2.22$ & $0.056$ & $1.83$ & 0.141 \\

			Rodi\cite{rodi1997comparison} & TL $k-\epsilon$ model & $1.25$ & $2.004$ & $0.07$ & $1.17$ & 0.143 \\

			Bosch\cite{bosch1998simulation} & TL $k-\epsilon$ model & $-$ & $1.750$ & $0.0012$ & $0.178$ & 0.122 \\		
			\lasthline
		\end{tabular}
		\begin{tablenotes}
			\item[a] TL $k-\epsilon$ model represents the two layer $k-\epsilon$ model.
		\end{tablenotes}
	\end{threeparttable}
\end{table}

The horizontal velocity distributed on the centerline is plotted in Fig.~\ref{figsqCylinderuc}, and the information of time averaged separation region behind the cylinder can be seen in the velocity profiles along the centerline. It shows a fairly well agreement in comparison to experimental and numerical approach data. Fig.~\ref{figsqCylinderu4} displays the streamwise velocity profiles at four positions behind a square cylinder. Very good agreement is obtained between present simulations and data extracted from literatures \cite{iaccarino2003reynolds}\cite{lyn1995laser}.

A qualitative picture of the vortex shedding behind the square cylinder is presented in Fig.~\ref{figsqCylindeStreamline4}. The streamlines are laid over on the Mach number contour plots. As expected, the alternative vortex shedding from upper and bottom side of the cylinder is shown clearly in the picture, and the vortexes are converted downstream in the wake of the cylinder.
\begin{table}[!htp]
	\centering
	\caption{\label{tableSqCylinderWorks} The comparison of computational works between explicit scheme and dual
    time-stepping strategy for the simulation of a square cylinder.}
	\begin{threeparttable}
        \begin{tabular}{p{90pt} p{40pt}<{\centering} p{80pt}<{\centering} p{70pt}<{\centering}}
			\hline
			\hline
            Scheme & $\Delta t$ & Inner iteration & Pseudo steady resudial\\
			\hline
            Explicit & $0.0006$ & $-$ & $-$ \\

            Dual time-stepping & $0.15$ & $20$ & $< 10^{-7}$ \\
			\lasthline
		\end{tabular}
	\end{threeparttable}
\end{table}

For the approach of incompressible turbulent flow around a square cylinder, the explicit time step can be obtained by the Eq.~\ref{exDeltaT}. The details of explicit time step and the physical time step of dual time-stepping strategy are shown in Table \ref{tableSqCylinderWorks}. It is evident that to predict the flow state at certain time $t$, the dual time-stepping method only costs about one-tenth of the computational works of the explicit scheme. The accuracy of approach is also guaranteed by using $20$ inner iterations in a single physical time step.

\subsection{Case 3: Transonic buffet on a NACA0012 airfoil}
For the transonic flow around an airfoil with certain combined conditions such as Mach number, Reynolds number, the airfoil profile and the angle of attack, a strong shock wave oscillations, which is termed as buffet, may be aroused and self-sustained even in the absence of any airfoil motion. Such a case studied in our paper is a transonic turbulent flow over a NACA0012 airfoil.  The Mach number of the free stream flow is $Ma_\infty = 0.72$. The Reynolds number, which is defined as $Re_\infty = \rho_\infty U_\infty c / \mu_\infty$, equals to $1\times 10^7$, where $c=1.0$ represents the chord length of NACA0012 airfoil. The angle of attack is $\alpha = 6^\circ$.

The aim of the case in this section is to validate the dual time-stepping method of gas-kinetic scheme in the simulation of unsteady transonic turbulent flow. For turbulent flow simulations, the Spalart-Allmaras (SA) turbulence model \cite{spalart1992one} is combined with the gas-kinetic scheme in present method. Actually the gas-kinetic scheme has been coupled with different types of turbulence models by some researchers.

The SA turbulence model is one of the popular models which is very suitable for simulation of separated flow. The details of the coupled methods \cite{FLD:FLD4239} and the turbulence model are not the focus of attentions in this study, which will not be described in detail here.

\begin{figure}[!htp]
	\centering
	\subfigure[Full domain]{
		\includegraphics[width=0.45 \textwidth]{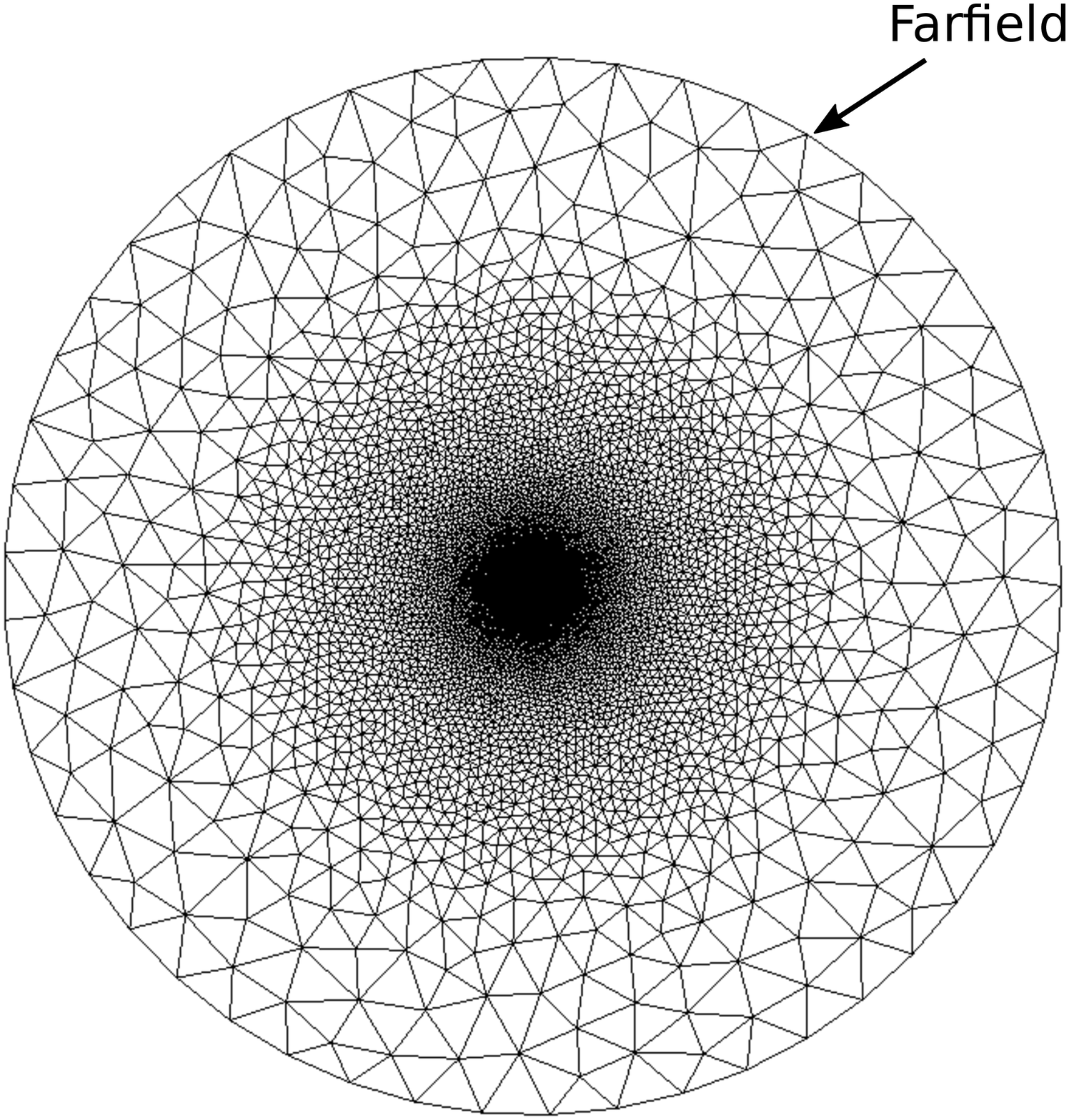}
		\label{fig0012Grid1}
	} \\
	\subfigure[View of grids near the airfoil]{
		\includegraphics[width=0.45 \textwidth]{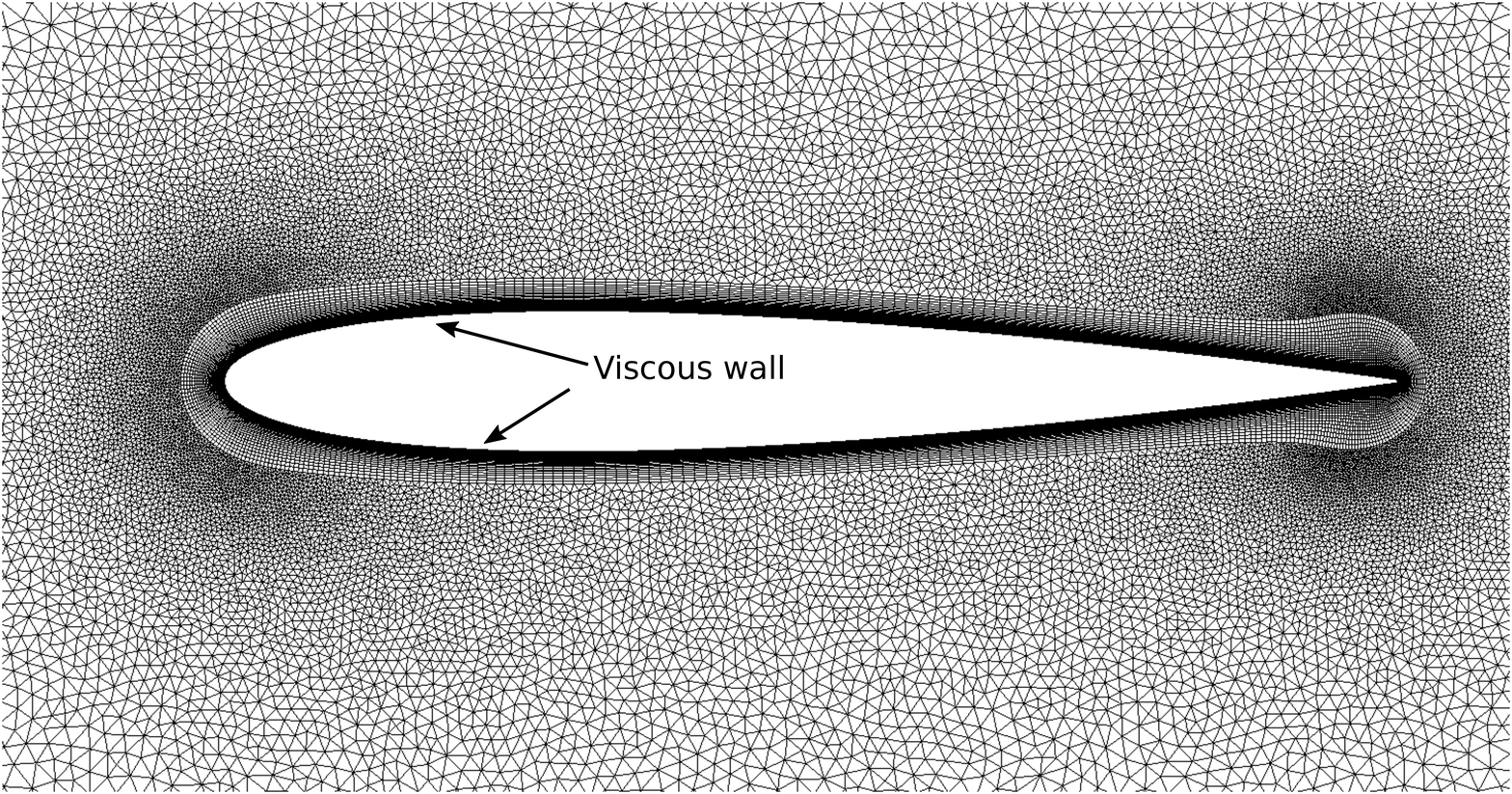}
		\label{fig0012Grid2}
	}
	\caption{Computational grids for NACA0012 airfoil.}
	\label{fig0012Grid}
\end{figure}

In Fig.~\ref{fig0012Grid}, the computational domain and the hybrid grids used in this approach are displayed. The total number of cells in the domain is $86665$. The rectangular meshes are used to maintain enough accuracy of numerical simulations of the flows within the boundary layer near the airfoil. The nearest distance of mesh points to the wall of airfoil is $2.5 \times 10^{-6}$, and the y-plus is $y^+ \approx 0.9$. The rectangle region is extruded $75$ layers from the wall of airfoil, and there are $400$ points located on the airfoil. The outer domain is about $50$ times chord length of the airfoil.

\begin{table}[!htp]
	\centering
	\caption{\label{tableNACA0012ExpSet}NACA0012 transonic buffet sets \cite{mcdevitt1985static}.}
	\begin{threeparttable}
		\begin{tabular}{p{65pt} p{25pt}<{\centering} p{100pt}<{\centering} p{100pt}<{\centering}}
			\hline
			\hline
			Set & $\alpha_\infty (^\circ)$ & $Ma_\infty$ & {$\overline{f}_{exp}$}\\
			\hline
			$6$ & $6$ & $0.72$ & $0.55$ \\
			
			$1$ & $4$ & $0.75$ & $0.47$  \\
			
			$5$ & $4$ & $0.77$ & $0.44$  \\
			
			$4$ & $4$ & $0.80$ & $0.38$  \\			
			\lasthline
		\end{tabular}
		\begin{tablenotes}
			\item[a] The Reynolds number of free stream flow in the experiments is about $Re_\infty \approx 1.0\times10^7$.
		\end{tablenotes}
	\end{threeparttable}
\end{table}

The experiment of NACA0012 transonic buffet was carried out by McDevitt and Okuno\cite{mcdevitt1985static} at the NASA Ames Research Center's high-Reynolds number facility. Four conditions sets which McDevitt and Okuno chose to obtain the stable self-sustained transonic buffet in the experiments are listed in Table \ref{tableNACA0012ExpSet}. $\overline{f}$
denotes the reduced frequency, which is defined as
\begin{equation}\label{reducedFrequency}
	\overline{f} = \frac{2{\pi}fc}{U_\infty}.
\end{equation}

In our study, the conditions of set 6 is chose for the test of transonic buffet on NACA0012 airfoil. To validate the current computational setup, the results of computed transonic buffet in our paper are compared with experiments and other numerical methods. Table \ref{tableNACA0012Clf} lists the details of comparisons using the conditions of set 6 in the reference \cite{mcdevitt1985static}. $\Delta C_l$ represents the amplitude of the lift coefficient, and $\Delta X$ denotes the distance of shock-buffet traveling on the airfoil surface. The results demonstrate a very good accordance with the references.
\begin{table}[!htp]
	\centering
	\caption{\label{tableNACA0012Clf} Transonic buffet frequency and the amplitude of lift coefficient.}
	\begin{threeparttable}
		\begin{tabular}{p{65pt} p{25pt}<{\centering} p{100pt}<{\centering} p{100pt} < {\centering} p{100pt}}
			\hline
			\hline
			  & Present & McDevitt\cite{mcdevitt1985static} &Iovnovich\cite{iovnovich2012reynolds} \\
			\hline
			$\overline{f}$ & $0.4879$ & $0.55$ &$0.5$ \\
			
			$\Delta C_l$& $0.41$ &$-$& $0.46$   \\
			
			$\Delta X/c$&$0.25$&$-$& $0.26$\\
			\lasthline
		\end{tabular}
	\end{threeparttable}
\end{table}

\begin{figure}[!htp]
	\centering
	\subfigure[$t=0$]{
		\includegraphics[width=0.4\textwidth]{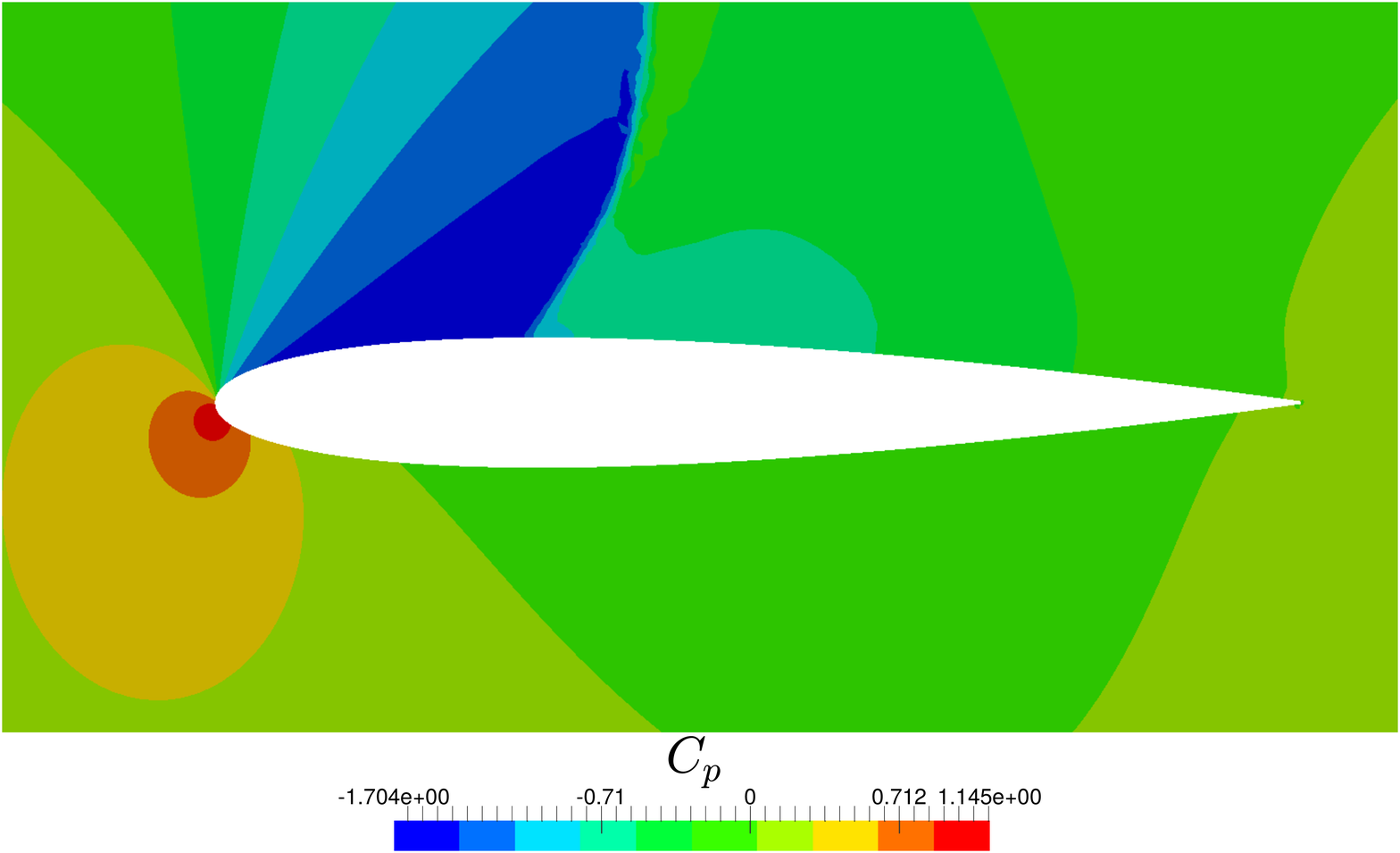}
		\label{figNACA0012Cp0}
	}
	\subfigure[$t=\frac{T}{4}$]{
		\includegraphics[width=0.4\textwidth]{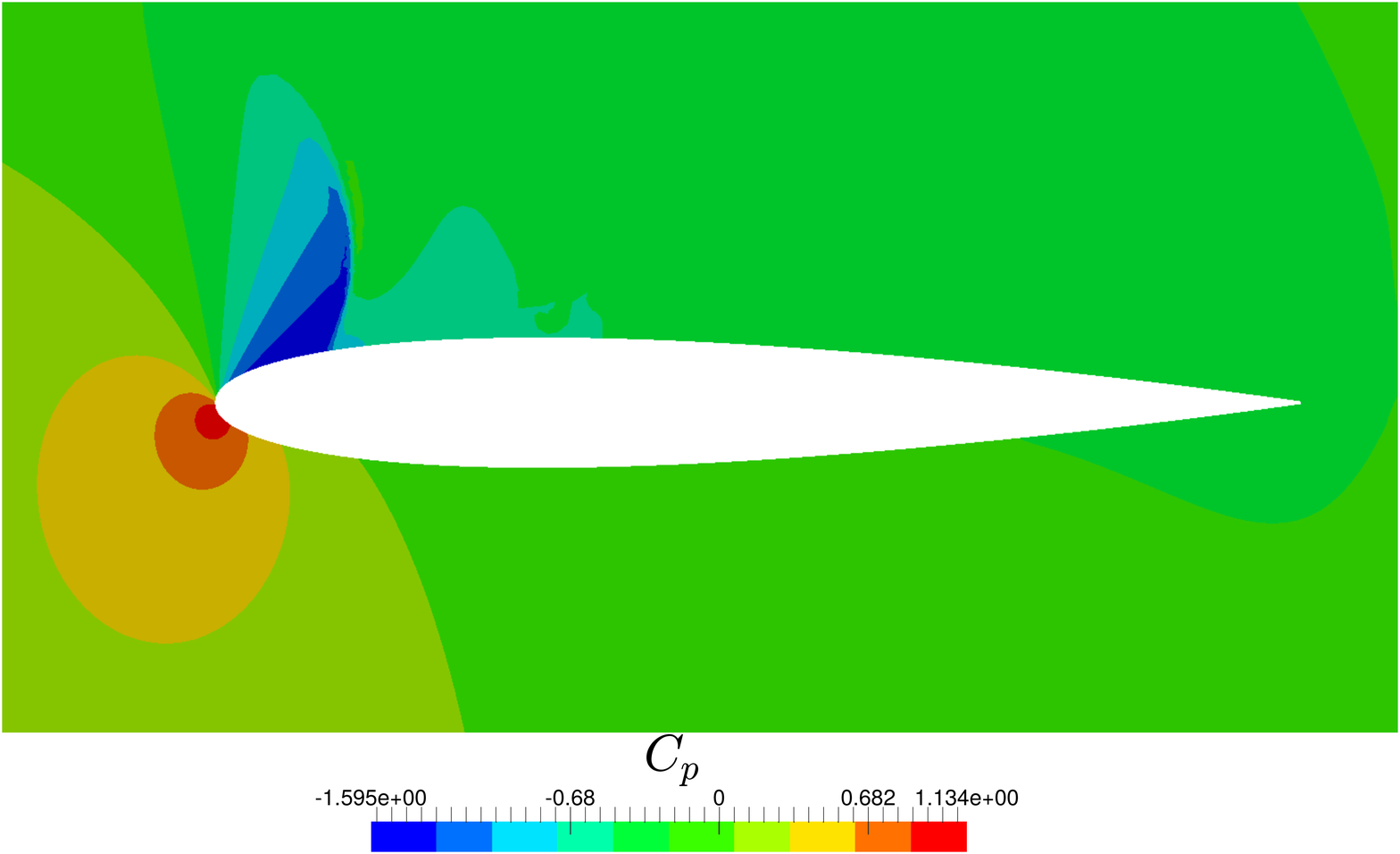}
		\label{figNACA0012Cp1}
	}
	\subfigure[$t=\frac{T}{2}$]{
		\includegraphics[width=0.4\textwidth]{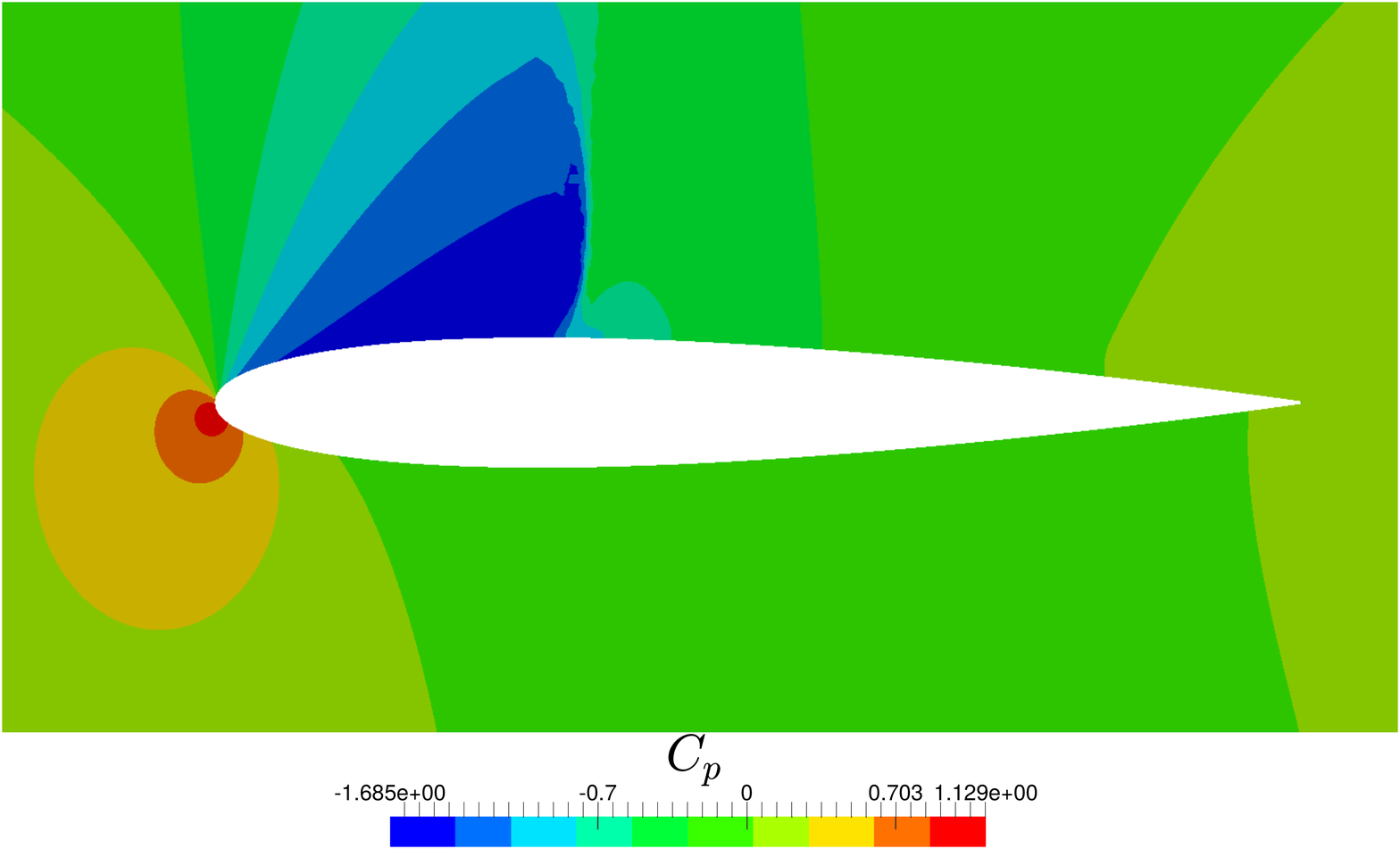}
		\label{figNACA0012Cp2}
	}
	\subfigure[$t=\frac{3T}{4}$]{
		\includegraphics[width=0.4\textwidth]{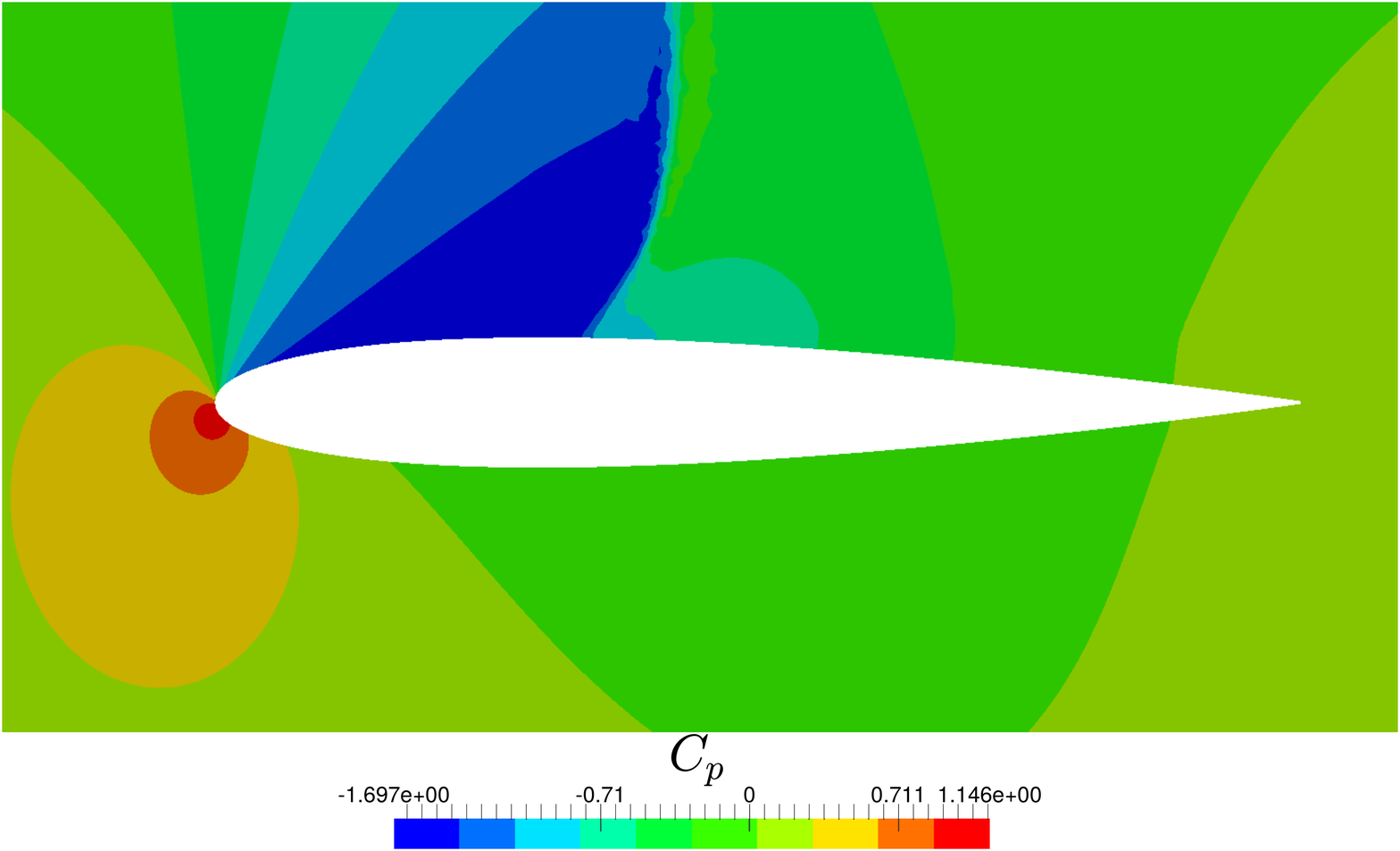}
		\label{figNACA0012Cp3}
	}
	\caption{The evolution of pressure coefficient in a period. T represents the period.}
	\label{figNACA0012Cp}
\end{figure}

\begin{figure}[!htp]
	\centering
	\includegraphics[width=0.4 \textwidth]{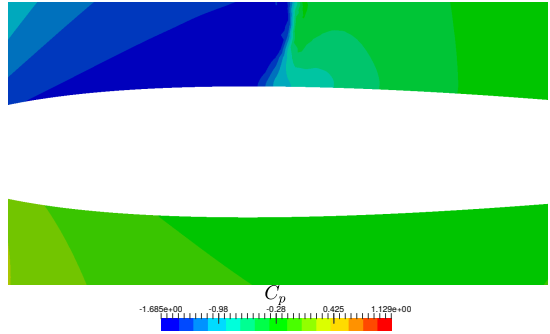}
	\caption{\label{figNACA0012SBLI} The  lambda-shock structure over a NACA0012 airfoil.}
\end{figure}
The evolution of the pressure coefficient in a period is plotted in Fig.~\ref{figNACA0012Cp}. As expected, the shuttle of shock-buffet is shown in this figure. Fig.~\ref{figNACA0012SBLI} displays the captured shock-wave boundary layer interaction. The lambda-shock structure can be seen in the plot. Since the resolution of mesh is insufficient for the flow at high Reynolds number, the lambda region is not well resolved and the lambda structure is not very clear.
\begin{figure}[!htp]
	\centering
	\includegraphics[width=0.4 \textwidth]{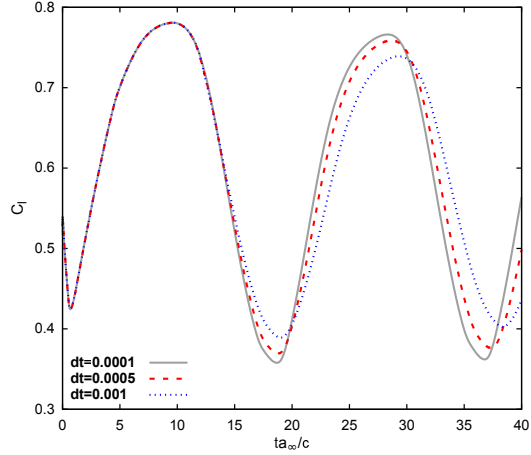}
	\caption{\label{figNACA0012Cl} Unsteady response to the physical time step. $a_\infty$ represents the sound speed of the free stream flow.}
\end{figure}

To study the effect of physical time step in the simulation on transonic buffet responses, simulations were performed using the physical time step ranging from $0.0001$ to $0.001$. Fig.~\ref{figNACA0012Cl} shows the time histories of lift coefficient at different time steps. The convergence is evident with decreasing physical time step, and the time step $0.0001$ is chose in our tests.
\begin{table}[!htp]
	\centering
	\caption{\label{tableNACA0012Works} The comparison of computational works between explicit scheme and dual
    time-stepping strategy for the simulation of transonic buffet on the NACA0012 airfoil.}
	\begin{threeparttable}
        \begin{tabular}{p{90pt} p{40pt}<{\centering} p{80pt}<{\centering} p{70pt}<{\centering}}
			\hline
			\hline
            Scheme & $\Delta t$ & Inner iteration & Pseudo steady resudial\\
			\hline
            Explicit & $6\times 10^{-7}$ & $-$ & $-$ \\

            Dual time-stepping & $0.0001$ & $10$ & $< 10^{-7}$ \\
			\lasthline
		\end{tabular}
	\end{threeparttable}
\end{table}

The computational works of explicit scheme and dual time-stepping method are also compared in Table
\ref{tableNACA0012Works}. From the Table \ref{tableNACA0012Works}, we can easily conclude that the dual time-stepping method can not only reduce the computational costs greatly, but also predict the transonic buffet with sufficient accuracy.

\section{Conclusions}
In present work, the dual time-stepping strategy of gas-kinetic scheme is proposed for the prediction of unsteady flows. The test cases not only cover viscous flows throughout the Mach number range from incompressible through transonic flows, but also cover the flows throughout the Reynolds number range from laminar to turbulent flows. All the three tests obtain a good agreement with the referred data and meet the goals which are designed for the validation. To accelerate the convergence of pseudo steady state, implicit gas-kinetic scheme is employed in the inner iteration. Both inviscid flux Jacobian and viscous flux Jacobian are considered in the construction of linear system, and GMRES method is adopted to approach the solution of linear system. It can be obviously seen in present study that the ability of dual time-stepping method to save the computational work is evident compared with explicit scheme. The good results demonstrate that the dual time-stepping strategy of gas-kinetic scheme can simulate unsteady flows accurately and efficiently, and the present work is of particular usefulness for unsteady flow predictions in the field of engineering.

\section*{Acknowledgements}
\begin{acknowledgments}
The project financially supported by National Natural Science Foundation of China (Grant No. 11472219), Natural Science Basic Research Plan in Shaanxi Province of China (Program No. 2015JM1002), as well as National Pre-Research Foundation of China.
\end{acknowledgments}

 \newcommand{\noop}[1]{}

\end{document}